\documentclass[preprint,showpacs,preprintnumbers,amsmath,amssymb,showkeys]{revtex4}

\usepackage{graphicx}
\usepackage{dcolumn}
\usepackage{bm}
\usepackage{braket}
\usepackage{enumitem}
\usepackage[utf8]{inputenc}
\usepackage{amsmath}
\usepackage{amssymb}
\usepackage{amsthm}

\begin{document}

\noindent

\preprint{}

\title{General quantum correlation from nonreal values of Kirkwood-Dirac quasiprobability over orthonormal product bases}

\author{Agung Budiyono$^{1,2,3}$}
\email{agungbymlati@gmail.com}
\author{Bobby E. Gunara$^4$}
\author{Bagus E. B. Nurhandoko$^{4}$}
\author{Hermawan K. Dipojono$^{2,3}$}
\affiliation{$^1$Research Center for Quantum Physics, National Research and Innovation Agency, South Tangerang 15314, Indonesia} 
\affiliation{$^2$Research Center for Nanoscience and Nanotechnology, Bandung Institute of Technology, Bandung, 40132, Indonesia}
\affiliation{$^3$Department of Engineering Physics, Bandung Institute of Technology, Bandung, 40132, Indonesia} 
\affiliation{$^4$Department of Physics, Bandung Institute of Technology, Bandung, 40132, Indonesia}

\date{\today}

\begin{abstract} 

We propose a characterization and a quantification of general quantum correlation which is exhibited even by a separable (unentangled) mixed bipartite state in terms of the nonclassical values of the associated Kirkwood-Dirac (KD) quasiprobability. Such a general quantum correlation, wherein entanglement is a subset, is not only intriguing from a fundamental point of view, but it has also been recognized as a resource in a variety of schemes of quantum information processing and quantum technology. Given a bipartite state, we construct a quantity based on the imaginary part the associated KD quasiprobability defined over a pair of orthonormal product bases and an optimization procedure over all pairs of such bases. We show that it satisfies certain requirements expected for a quantifier of general quantum correlations. It gives a lower bound to the total sum of the quantum standard deviation of all the elements of the product (local) basis, minimized over all such bases. It suggests an interpretation as the minimum genuine quantum share of uncertainty in all possible local von-Neumann projective measurement. Moreover, it is a faithful witness for entanglement and measurement-induced nonlocality of pure bipartite states. We then discuss a variational scheme for its estimation, and based on this, we offer information theoretical meanings of the general quantum correlation. Our results suggest a deep connection between the general quantum correlation and the nonclassical values of the KD quasiprobability and the associated strange weak values.  

\end{abstract} 

\maketitle       

\section{Introduction}

Nonclassical correlation in multipartite quantum systems was originally associated with entangled states, namely states that cannot be prepared by any set of local operation and classical communication (LOCC) \cite{Horodecki entanglement review}. Studies in the last couple of decades however showed that most unentangled (i.e., separable) but mixed states are somewhat nonclassically correlated, manifested in the various forms of discord-like quantum correlations \cite{Ollivier quantum discord paper,Henderson quantum discord paper,Oppenheim work deficit,Horodecki work deficit 1,Luo measurement induced disturbance approach to quantum correlation: quantum deficit,Dakic necessary and sufficient condition for geometric quantum discord,Modi geometric discord,Luo geometric discord,Monras geometric discord,Piani geometric discord,Luo measurement induced nonlocality,Piani discord generates entanglement via measurement,Streltsov discord generates entanglement,Gharibian discord generates entanglement,Giampaolo global change due to local unitary,Gharibian global change due to local unitary,Seshadreesan global change induced by entanglement breaking channel,Girolami quantum correlation based on WY skew information,Girolami quantum correlation based on Fisher information,Mahdian dynamics of discord three qubits}. The nonclassical correlations beyond entanglement have received a growing attention both from a fundamental view point and also practically to better understand the physical origin of the quantum advantages in certain information processing tasks and schemes of quantum technology which consume very little or no entanglement \cite{Datta discord and one qubit quantum computation,Lanyon discord and quantum computation,Passante discord and quantum computation,Piani discord and no-local broadcasting,Dakic discord and remote state preparation,Cavalcanti discord and quantum communication,Gu and discord quantum communication,Modi discord and quantum metrology}. Different approaches have been proposed to characterize and quantify the general quantum correlation, adopting ideas from quantum information theory within the framework of quantum resource theory. See Refs. \cite{Modi discord and the boundary of quantum and classical correlation review,Adesso quantum correlation beyond entanglement review,Bera quantum discord review} for recent reviews. Conceptually, these general quantum correlations arise from the noncommutativity between the multipartite quantum state and some set of local quantum observables, rather than from the nonseparability of the state. 

On the other hand, there is an informationally equivalent representation of quantum state based on quasiprobability distributions where quantum noncommutativity appears as a different form of nonclassicality. Quasiprobability distributions are the quantum analogs of phase space probability distribution in classical statistical mechanics. Because of the quantum noncommutativity, they necessarily do not satisfy all the Kolmogorov axioms for probability. In particular, in a specific quasiprobability distribution called Kirkwood-Dirac (KD) quasiprobability \cite{Kirkwood quasiprobability,Dirac quasiprobability,Barut KD quasiprobability,Chaturvedi KD quasiprobability}, the failure of commutativity manifests in the nonvanishing imaginary part and/or negative values of its real part. Remarkably, the nonreality and/or the negativity of the KD quasiprobability is tighter than noncommutativity \cite{Drori nonclassicality tighter and noncommutativity,deBievre nonclassicality in KD distribution}. Significant works over the past decade showed that KD quasiprobability and its nonclassical values play crucial roles in quantum tomography \cite{Salvail direct measurement KD distribution,Bamber measurement of KD distribution,Lundeen measurement of KD distribution,Thekkadath measurement of density matrix}, quantum metrology \cite{Arvidsson-Shukur quantum advantage in postselected metrology,Lostaglio contextuality in quantum linear response,Lupu-Gladstein negativity enhanced quantum phase estimation 2022}, quantum thermodynamics  \cite{Lostaglio contextuality in quantum linear response, Allahverdyan TMH as quasiprobability distribution of work,Lostaglio TMH quasiprobability fluctuation theorem contextuality}, a wide spectrum of quantum fluctuations in condensed matter physics \cite{Lostaglio KD quasiprobability and quantum fluctuation}, information scrambling in many body quantum chaos \cite{Alonso KD quasiprobability witnesses quantum scrambling,Halpern quasiprobability and information scrambling}, in quantum foundation to prove quantum contextuality \cite{Pusey negative TMH quasiprobability and contextuality,Kunjwal KD quasiprobability and contextuality}, and recently in the characterization and quantificiation of coherence and asymmetry \cite{Agung KD-nonreality coherence,Agung translational asymmetry from nonreal weak value,Agung Budiyono characterization of trace-norm asymmetry in terms of nonreal weak values}.    

In this paper, we attempt to develop a relation between the above two notions of nonclassicality which both arise directly from noncommutativity, i.e., the general quantum correlation in a multipartite state and the nonclassical values of the associated KD quasiprobability. Namely, we ask: can we use the nonclassical values of the KD quasiprobability to characterize and quantify the general quantum correlation? First, given a multipartite quantum state, we construct a quantity from the imaginary part of the KD quasiprobability defined over a pair of orthonormal product bases of the relevant Hilbert space, and optimized over all possible choices of such bases. We argue that it satisfies certain plausible requirements expected for a quantifier of quantum correlation beyond entanglement, and thus refer to it as KD nonclassical correlation. KD nonclassical correlation sets a lower bound to the uncertainty arising in von-Neumann local projective measurement, quantified by the total sum of the quantum standard deviation of all the elements of the associated local PVM (projection-valued measure), minimized over all such possible projective measurements. It may be interpreted as the minimum genuine quantum share out of the total uncertainty arising in any measurement described by local PVM over the multipartite state. For arbitrary pure bipartite states, it is a faithful witness for the linear entropy of entanglement and measurement-induced nonlocality \cite{Luo measurement induced nonlocality}. We then discuss a scheme to estimate the KD nonclassical correlation directly without recoursing to quantum state tomography, via a hybrid quantum-classical variational circuit by combining a reconstruction of the KD quasiprobability or the associated weak value \cite{Aharonov weak value,Aharonov-Daniel book,Wiseman weak value,Tamir weak value review,Johansen quantum state from successive projective measurement,Johansen weak value from a sequence of strong measurement,Salvail direct measurement KD distribution,Bamber measurement of KD distribution,Lundeen measurement of KD distribution,Thekkadath measurement of density matrix,Lostaglio KD quasiprobability and quantum fluctuation,Jozsa complex weak value,Haapasalo generalized weak value,Cohen estimating of weak value with strong measurements,Vallone strong measurement to reconstruct quantum wave function,Wagner measuring weak values and KD quasiprobability} and a classical optimization procedure. The scheme suggests new information theoretical meanings of the general quantum correlation. 

The article is organized as follows. In Section \ref{Nonclassical state and noncommutativity}, we summarize the notion of nonclassical multipartite states based on the noncommutativity between the multipartite state and some local basis. In Section \ref{Kirkwood-Dirac quasiprobability over product bases}, we define a specific class of KD quasiprobability over a pair of orthonormal product bases, and discuss its properties. In Section \ref{General nonclassical correlation from nonreal Kirkwood-Dirac quasiprobability}, we define KD nonclassical correlation in a bipartite quantum state based on the imaginary part of the associated KD quasiprobability over a pair of orthonormal product bases and optimization over all pairs of such bases. We show that it satisfies certain plausible requirements for a quantifier of general quantum correlation, and discuss its relation with negativity of quantumness. We also give numerical calculations of the KD nonclassical correlation in a maximally entangled two-qubit state and a $2\times 2$ Werner state. We proceed in Section \ref{KD nonclassical correlation as the witness for genuine local quantum uncertainty and pure state entanglement} to study the relation between the KD nonclassical correlation with the uncertainty arising in the measurement described by local PVM, and the linear entropy of entanglement and measurement-induced nonlocality for pure states. In Section \ref{Observation of KD nonclassical correlation and its information theoretical meaning}, we discuss a variational scheme for a direct estimation of the KD nonclassical correlation and suggest its information theoretical meaning. Section \ref{Summary and Remarks} ends with a summary and a remark. 

\section{Preliminaries}

\subsection{Nonclassical state and noncommutativity \label{Nonclassical state and noncommutativity}}

Consider a composite of two subsystems $AB$ with a quantum state that is represented by a density operator $\varrho_{AB}$ acting on a Hilbert space $\mathcal{H}_{AB}=\mathcal{H}_A\otimes\mathcal{H}_B$, where $\mathcal{H}_{A(B)}$ is the Hilbert space of the subsystem $A(B)$. In a general multipartite setting, a separable state is intuitively defined as a state that can be prepared by LOCC \cite{Horodecki entanglement review}. For a bipartite system, a separable state can thus be in general expressed as a classical statistical mixture of the product states, i.e., $\varrho_{AB}^{\mathcal{S}}=\sum_kp_k\varrho_A^k\otimes\varrho_B^k$, where $\{\varrho_{A(B)}^k\}$ is a set of density operators of the subsystem $A(B)$ on $\mathcal{H}_{A(B)}$, and $\{p_k\}$ are mixing probabilities: $p_k\ge 0$, $\sum_kp_k=1$. All other states, i.e., nonseparable states, are entangled. For pure states, separability is equivalent to the absence of nonclassical correlation. Remarkably, allowing impurity to the bipartite state may lead to nonclassical correlations even when it is separable. That is, various different schemes have been revealed over the past couple of decades which show that mixed separable states may nevertheless exhibit correlation that cannot be accessed by any classical object \cite{Modi discord and the boundary of quantum and classical correlation review,Adesso quantum correlation beyond entanglement review,Bera quantum discord review,Ollivier quantum discord paper,Henderson quantum discord paper,Oppenheim work deficit,Horodecki work deficit 1,Luo measurement induced disturbance approach to quantum correlation: quantum deficit,Dakic necessary and sufficient condition for geometric quantum discord,Modi geometric discord,Luo geometric discord,Monras geometric discord,Piani geometric discord,Luo measurement induced nonlocality,Piani discord generates entanglement via measurement,Streltsov discord generates entanglement,Gharibian discord generates entanglement,Giampaolo global change due to local unitary,Gharibian global change due to local unitary,Seshadreesan global change induced by entanglement breaking channel,Girolami quantum correlation based on WY skew information,Girolami quantum correlation based on Fisher information}. For instance, any local measurement to a separable but mixed multipartite state may yield a disturbance to the global state which cannot be locally accounted for \cite{Henderson quantum discord paper,Ollivier quantum discord paper,Luo measurement induced nonlocality}. 

By contrast, classical intuition suggests that for a genuine classical multipartite state, there is at least a local measurement which does not lead to a modification to the global state. For a bipartite system $AB$, there are two classes of states which conform with the above intuition \cite{Luo measurement induced disturbance approach to quantum correlation: quantum deficit,Piani discord and no-local broadcasting}. One is the class of classical-quantum (or, $A-$classically correlated) states, which are those that can be expressed as 
\begin{eqnarray}
\varrho_{AB}^{\mathcal CQ}:=\sum_k p_k\ket{k}\bra{k}_A\otimes\varrho_B^k, 
\label{classical-quantum state}
\end{eqnarray}
where $\{\ket{k}_A\}$ is an orthonormal basis of the Hilbert space $\mathcal{H}_A$ of  subsystem $A$, $\{\varrho_B^k\}$ is a set of states of subsystem $B$, and $\{p_k\}$ are probabilities: $p_k\ge 0$, $\sum_kp_k=1$. The quantum-classical (or, $B$-classically correlated) state is defined analogously by exchanging the role of the parties $A$ and $B$. The other class consists of bipartite states having the following form:
\begin{eqnarray}
\varrho_{AB}^{\mathcal CC}:=\sum_{k,l} p_{kl}\ket{k}\bra{k}_A\otimes\ket{l}\bra{l}_B, 
\label{classical-classical state}
\end{eqnarray}
where $\{\ket{k}_A\}$ and $\{\ket{l}_B\}$ are orthonormal bases of subsystem $A$ and $B$, respectively, and $\{p_{kl}\}$ are joint probabilities: $p_{kl}\ge 0$, $\sum_{k,l}p_{kl}=1$. Such states are called classical-classical (or, totally classically correlated) states. One can see that applying local projective measurements described by PVMs $\{\Pi_A^k\otimes\mathbb{I}_B\}$ and $\{\Pi_A^k\otimes\Pi_B^l\}$, where $\Pi^x:=\ket{x}\bra{x}$, respectively, to the classical-quantum states and classical-classical states, without learning the outcomes (non-selective measurement), leave them unmodified, i.e., $\sum_k(\Pi_A^k\otimes\mathbb{I}_B)\varrho_{AB}^{\mathcal CQ}(\Pi_A^k\otimes\mathbb{I}_B)=\varrho_{AB}^{\mathcal CQ}$ and $\sum_{k,l}(\Pi_A^k\otimes\Pi_B^l)\varrho_{AB}^{\mathcal CC}(\Pi_A^k\otimes\Pi_B^l)=\varrho_{AB}^{\mathcal CC}$. This can also be seen as due to the fact that the local measurement bases (i.e., the PVMs) and the bipartite state commute: $[(\Pi_A^k\otimes\mathbb{I}_B),\varrho_{AB}^{\mathcal CQ}]=0$ and $[(\Pi_A^k\otimes\Pi_B^l),\varrho_{AB}^{\mathcal CC}]=0$, for all $k$, $l$. Any bipartite state that cannot be represented either in the forms of Eqs. (\ref{classical-quantum state}) or (\ref{classical-classical state}) has been shown to contain some kinds of nonclassical correlation arising from the noncommutativity between the bipartite state and the local measurement basis, even if it is separable \cite{Adesso quantum correlation beyond entanglement review}.  

Various mathematical characterizations of the classical bipartite states of the types of Eqs. (\ref{classical-quantum state}) and (\ref{classical-classical state})  have led researchers to develop different quantifiers and measures of quantum correlation beyond entanglement in a bipartite state, by applying quantum information theoretical concepts: quantum mutual information, von-Neumann entropy, distance and infidelity between two density operators, Wigner-Yanase skew information and quantum Fisher information \cite{Adesso quantum correlation beyond entanglement review}. All these quantifiers essentially quantify the difference between the bipartite state under scrutiny and the class of classical bipartite states of Eqs. (\ref{classical-quantum state}) and (\ref{classical-classical state}). They therefore directly or indirectly capture the failure of commutativity between the bipartite state and some set of local observables. In order to better understand the operational and statistical meaning of the quantum correlation beyond entanglement, it is desirable to have a quantifier of general quantum correlation whose definition directly and transparently translates into laboratory operations. Moreover, it is also important to study general quantum correlation from as many angles as possible to reveal its rich facets.  For these reasons, we shall add to the above list with a characterization and a quantification of the general quantum correlation based on the nonclassical values of KD quasiprobability which admits direct interpretation in terms of laboratory operations.    

\subsection{Kirkwood-Dirac quasiprobability over orthonormal product bases \label{Kirkwood-Dirac quasiprobability over product bases}}

KD quasiprobability is an equivalent representation of quantum state suited for finite dimensional Hilbert space, in which quantum noncommutativity implies that its value is not necessarily real and nonnegative \cite{Kirkwood quasiprobability,Dirac quasiprobability,Barut KD quasiprobability,Chaturvedi KD quasiprobability}. Consider again a bipartite quantum system $AB$. We first construct a pair of orthonormal bases of the Hilbert space $\mathcal{H}_{AB}$ of the bipartite system by taking the tensor product of the orthonormal basis of each subsystem as: $\{\ket{a_1,b_1}_{AB}:=\ket{a_1}_A\otimes\ket{b_1}_B\}$ and $\{\ket{a_2,b_2}_{AB}:=\ket{a_2}_A\otimes\ket{b_2}_B\}$, where $\{\ket{a_1}_A\}$ and $\{\ket{a_2}_A\}$ are two orthonormal bases of $\mathcal{H}_A$ of subsystem $A$, and $\{\ket{b_1}_B\}$ and $\{\ket{b_2}_B\}$ are two orthonormal bases of $\mathcal{H}_B$ of subsystem $B$. From here on, the Roman letter in the element of the basis denotes which subsystem (i.e., subsystem $A$, subsystem $B$, etc.), while the Arabic number denotes which basis (i.e., the first basis, the second basis, etc.) of the Hilbert space of each subsystem. For example, $\{\ket{a_1}\}$ is the first orthonormal basis of the Hilbert space $\mathcal{H}_A$ of subsystem $A$, etc. \\
{\bf Definition 1}. The KD quasiprobability associated with a bipartite density operator $\varrho_{AB}$ on a finite-dimensional Hilbert space $\mathcal{H}_{AB}$ over a pair of orthonormal product bases $\{\ket{a_1,b_1}\}$ and $\{\ket{a_2,b_2}\}$ of $\mathcal{H}_{AB}$ is defined as   
\begin{eqnarray}
{\rm Pr}_{\rm KD}(a_1,b_1;a_2,b_2|\varrho_{AB}):={\rm Tr}\big((\Pi_A^{a_2}\otimes\Pi_B^{b_2})(\Pi_A^{a_1}\otimes\Pi_B^{b_1})\varrho_{AB}\big). 
\label{KD quasiprobability 1}
\end{eqnarray}

The (marginal) KD quasiprobability associated with a subsystem is obtained by summing over the variables corresponding to the other subsystem. For example, summing over $b_1,b_2$ of subsystem $B$, one gets the KD quasiprobability associated with subsystem $A$: $\sum_{b_1,b_2}{\rm Pr}_{\rm KD}(a_1,b_1;a_2,b_2|\varrho_{AB})={\rm Tr}((\Pi_A^{a_2}\otimes\mathbb{I}_B)(\Pi_A^{a_1}\otimes\mathbb{I}_B)\varrho_{AB})={\rm Tr}_A(\Pi_A^{a_1}\Pi_A^{a_2}\varrho_A):={\rm Pr}_{\rm KD}(a_1,a_2|\varrho_A)$, where $\varrho_A={\rm Tr}_B\varrho_{AB}$, and we have used the completeness relation: $\sum_{b_1}\Pi_B^{b_1}=\sum_{b_2}\Pi_B^{b_2}=\mathbb{I}_B$. Similarly, we have $\sum_{a_1,a_2}{\rm Pr}_{\rm KD}(a_1,b_1;a_2,b_2|\varrho_{AB})={\rm Tr}_B(\Pi_B^{b_1}\Pi_B^{b_2}\varrho_B):={\rm Pr}_{\rm KD}(b_1,b_2|\varrho_B)$, $\varrho_B={\rm Tr}_A\varrho_{AB}$. The above definition extends naturally to more than two parties. For example, for a tripartite quantum state $\varrho_{ABC}$ on a Hilbert space $\mathcal{H}_{ABC}=\mathcal{H}_A\otimes\mathcal{H}_B\otimes\mathcal{H}_C$, and a pair of orthonormal product bases $\{\ket{a_1,b_1,c_1}_{ABC}\}$ and $\{\ket{a_2,b_2,c_2}_{ABC}\}$ of $\mathcal{H}_{ABC}$, where $\{\ket{x_i}\}$, $i=1,2$, is the $i$th orthonormal basis of the Hilbert space $\mathcal{H}_X$ of subsystem $X$, $X=A,B,C$, we define the associated KD quasiprobability over the pair of the orthonormal product bases as
\begin{eqnarray}
{\rm Pr}_{\rm KD}(a_1,b_1,c_1;a_2,b_2,c_2|\varrho_{ABC}):={\rm Tr}\big((\Pi_A^{a_2}\otimes\Pi_B^{b_2}\otimes\Pi_C^{c_2})(\Pi_A^{a_1}\otimes\Pi_B^{b_1}\otimes\Pi_C^{c_1})\varrho_{ABC}\big). 
\label{KD quasiprobability three party}
\end{eqnarray}

One can see that the KD quasiprobability gives correct marginal probabilities in the following sense: $\sum_{a_1,b_1}{\rm Pr}_{\rm KD}(a_1,b_1;a_2,b_2|\varrho_{AB})={\rm Tr}((\Pi_A^{a_2}\otimes\Pi_B^{b_2})\varrho_{AB})={\rm Pr}(a_2,b_2|\varrho_{AB})$, and $\sum_{a_2,b_2}{\rm Pr}_{\rm KD}(a_1,b_1;a_2,b_2|\varrho_{AB})={\rm Tr}((\Pi_A^{a_1}\otimes\Pi_B^{b_1})\varrho_{AB})={\rm Pr}(a_1,b_1|\varrho_{AB})$, where ${\rm Pr}(\cdot)$ denotes the usual real and non-negative classical probability. One also has single variable marginal probabilities, $\sum_{b_1,a_2,b_2}{\rm Pr}_{\rm KD}(a_1,b_1;a_2,b_2|\varrho_{AB})={\rm Tr}_A(\Pi_A^{a_1}\varrho_A)={\rm Pr}(a_1|\varrho_A)$, and $\sum_{a_1,a_2,b_2}{\rm Pr}_{\rm KD}(a_1,b_1;a_2,b_2|\varrho_{AB})={\rm Tr}_B(\Pi_B^{b_1}\varrho_B)={\rm Pr}(b_1|\varrho_B)$, etc. Hence, KD quasiprobability is normalized, i.e., $\sum_{a_1,b_1,a_2,b_2}{\rm Pr}_{\rm KD}(a_1,b_1;a_2,b_2|\varrho_{AB})=1$.  However and importantly, despite all of the above classically desirable properties, the KD quasiprobability ${\rm Pr}_{\rm KD}(a_1,b_1;a_2,b_2|\varrho_{AB})$ may take nonreal values and its real part may be negative. 

We also define the KD quasiprobability over three variables as follows which will play a central role in our characterization of general quantum correlation later. \\
{\bf Definition 2}. Summing the KD quasiprobability ${\rm Pr}_{\rm KD}(a_1,b_1;a_2,b_2|\varrho_{AB})$ over one of the variables we get KD quasiprobability over the three remaining variables.  For example, let us sum over $b_1$ to get 
\begin{eqnarray}
{\rm Pr}_{\rm KD}(a_1;a_2,b_2|\varrho_{AB})&:=&\sum_{b_1}{\rm Pr}_{\rm KD}(a_1,b_1;a_2,b_2|\varrho_{AB})\nonumber\\
&=&{\rm Tr}\big((\Pi_A^{a_2}\otimes\Pi_B^{b_2})(\Pi_A^{a_1}\otimes\mathbb{I}_B)\varrho_{AB}\big). 
\label{KD quasiprobability 2}
\end{eqnarray}

${\rm Pr}_{\rm KD}(a_1;a_2,b_2|\varrho_{AB})$ too has correct marginal probabilities, thus normalized, but may assume nonreal values and its real part may be negative. 

In this sense, the nonreality and/or the negativity of the KD quasiprobability associated with a quantum state, indicates a form of quantumness or nonclassicality, known as KD nonclassicality. It arises from the failure of commutativity among the state and the pair of defining bases. To see this within the KD quasiprobability over orthonormal product bases introduced above, consider for example the KD quasiprobability of Eq. (\ref{KD quasiprobability 1}) and assume that the first basis and the bipartite state commute, i.e., $[\Pi_A^{a_1}\otimes\Pi_B^{b_1},\varrho_{AB}]=0$, for all $a_1$ and $b_1$. Then, noting that $(\Pi_A^{a_1}\otimes\Pi_B^{b_1})^2=\Pi_A^{a_1}\otimes\Pi_B^{b_1}$, Eq. (\ref{KD quasiprobability 1}) can be written as ${\rm Pr}_{\rm KD}(a_1,b_1;a_2,b_2|\varrho_{AB})={\rm Tr}\big((\Pi_A^{a_2}\otimes\Pi_B^{b_2})\frac{(\Pi_A^{a_1}\otimes\Pi_B^{b_1})\varrho_{AB}(\Pi_A^{a_1}\otimes\Pi_B^{b_1})}{{\rm Tr}((\Pi_A^{a_1}\otimes\Pi_B^{b_1})\varrho_{AB})}\big){\rm Tr}((\Pi_A^{a_1}\otimes\Pi_B^{b_1})\varrho_{AB})$, which is just the probability to get $(a_1,b_1)$ and then subsequently $(a_2,b_2)$ in a sequence of measurements, so that it is real and nonnegative. Note in particular that the noncommutativity between the state and the orthonormal bases is formally directly captured by the imaginary part of the KD quasiprobability: ${\rm Im}({\rm Pr}_{\rm KD}(a_1,b_1;a_2,b_2|\varrho_{AB}))=(1/2i)\braket{a_2,b_2|[(\Pi_A^{a_1}\otimes\Pi_B^{b_1}),\varrho_{AB}]|a_2,b_2}$. As an example of KD quasiprobability with nonreal values, consider the following maximally entangled two-qubit state:
\begin{eqnarray}
\ket{\Psi_{AB}^{\rm me}}=\frac{1}{\sqrt{2}}\ket{00}+\frac{1}{\sqrt{2}}\ket{11}, 
\label{maximally entangled two-qubit state} 
\end{eqnarray} 
where $\{\ket{0},\ket{1}\}$ is the computational basis, i.e., the eigenstates of Pauli operator $\sigma_z$. Let us choose the following orthonormal product bases for the first and the second bases: $\{\ket{a_1,b_1}\}=\{\ket{0,0},\ket{0,1},\ket{1,0},\ket{1,1}\}$ and $\{\ket{a_2,b_2}\}=\{\ket{x_+,y_+},\ket{x_+,y_-},\ket{x_-,y_+},\ket{x_-,y_-}\}$, where $\ket{x_{\pm}}=\frac{1}{\sqrt{2}}(\ket{0}\pm\ket{1})$ and $\ket{y_{\pm}}=\frac{1}{\sqrt{2}}(\ket{0}\pm i\ket{1})$. Then, the associated KD quasiprobability in matrix expression with the first basis taking the row index and the second basis taking the column index, has the form:
\begin{eqnarray}
&&\{{\rm Pr}_{\rm KD}(\varrho)\}_{(a_1b_1,a_2b_2)}=
\begin{pmatrix} 
 \frac{1}{8}+\frac{1}{8}i & \frac{1}{8}-\frac{1}{8}i&  \frac{1}{8}-\frac{1}{8}i&  \frac{1}{8}+\frac{1}{8}i\\
0& 0& 0&  0\\
0& 0& 0&  0\\
\frac{1}{8}-\frac{1}{8}i & \frac{1}{8}+\frac{1}{8}i&  \frac{1}{8}+\frac{1}{8}i&  \frac{1}{8}-\frac{1}{8}i
\end{pmatrix}.
\end{eqnarray}
One can check that it has correct marginal probabilities, and normalized to unity. Moreover, the real parts are all nonnegative. Hence, the noncommutativity among the bipartite state and the pair of defining bases manifests entirely in the nonvanishing imaginary part. The above observation naturally raises the question on how the KD nonclassicality, and in particular, the imaginary part of the KD quasiprobabilities, are related to the general quantum correlation which also captures noncommutativity between the multipartite state and some local measurements. 

\section{General quantum correlation from the nonreality of Kirkwood-Dirac quasiprobability over orthonormal product bases\label{General nonclassical correlation from nonreal Kirkwood-Dirac quasiprobability}}

We wish to establish a quantitative link between the nonclassicality encoded in a bipartite quantum state $\varrho_{AB}$ captured by the concept of general quantum correlation and the nonclassical values in the associated KD quasiprobability ${\rm Pr}_{\rm KD}(a_1,b_2;a_2,b_2|\varrho_{AB})$ over a pair of orthonormal product bases. To this end, we will devise a quantifier of general quantum correlation in a bipartite state from the imaginary part of the KD quasiprobability. \\
{\bf Definition 3}. Given an arbitrary bipartite quantum state $\varrho_{AB}$ on a finite-dimensional Hilbert space $\mathcal{H}_{AB}$, we define the following quantity which maps the state to a nonnegative real number:
\begin{eqnarray}
&&\mathcal{Q}_{\rm KD}^A(\varrho_{AB})\nonumber\\
&:=&\inf_{\{\ket{a_1}\}}\sup_{\{\ket{a_2,b_2}\}}\sum_{a_1,a_2,b_2}\big|{\rm Im}{\rm Pr}_{\rm KD}(a_1;a_2,b_2|\varrho_{AB})\big|\nonumber\\
&=&\inf_{\{\ket{a_1}\}}\sup_{\{\ket{a_2,b_2}\}}\sum_{a_1,a_2,b_2}\big|{\rm Im}\braket{a_2,b_2|(\Pi_A^{a_1}\otimes\mathbb{I}_B)\varrho_{AB}|a_2,b_2}\big|\nonumber\\
&=&\inf_{\{\ket{a_1}\}}\sup_{\{\ket{a_2,b_2}\}}\sum_{a_1,a_2,b_2}\frac{1}{2}\big|\braket{a_2,b_2|[(\Pi_A^{a_1}\otimes\mathbb{I}_B),\varrho_{AB}]|a_2,b_2}\big|,
\label{one-sided KD nonclassical correlation} 
\end{eqnarray}
where the supremum is taken over the set $\mathcal{B}_{\rm op}(\mathcal{H}_{AB})$ of all orthonormal product bases of the Hilbert space $\mathcal{H}_{AB}$, and the infimum is taken over the set $\mathcal{B}_{\rm o}(\mathcal{H}_A)$ of all orthonormal bases of $\mathcal{H}_A$. 

Hence, $\mathcal{Q}_{\rm KD}^A(\varrho_{AB})$ is the total sum of the absolute imaginary part of the KD quasiprobability ${\rm Pr}_{\rm KD}(a_1;a_2,b_2|\varrho_{AB})$ defined in Eq. (\ref{KD quasiprobability 2}), first maximized over all orthonormal product bases $\{\ket{a_2,b_2}\}\in\mathcal{B}_{\rm op}(\mathcal{H}_{AB})$ of $\mathcal{H}_{AB}$, and then minimized over all local orthonormal bases $\{\ket{a_1}\}\in\mathcal{B}_{\rm o}(\mathcal{H}_A)$ of $\mathcal{H}_A$. $\mathcal{Q}_{\rm KD}^A(\varrho_{AB})$ defined in Eq. (\ref{one-sided KD nonclassical correlation}) is intended to quantify the general quantum correlation in the bipartite state $\varrho_{AB}$ arising from the noncommutativity between the bipartite state and any PVM $\{\Pi_A^{a}\otimes\mathbb{I}_B\}$ describing local projective measurement over the subsystem $A$. 

Similarly, to quantify the general quantum correlation in the bipartite state $\varrho_{AB}$ stemming from the noncommutativity between the bipartite state and any product PVM $\{\Pi_A^{a}\otimes\Pi_B^{b}\}$ describing local projective measurement on $A$ and $B$, we define the following quantity.\\
{\bf Definition 4}. Given a bipartite state $\varrho_{AB}$ on a finite-dimensional Hilbert space $\mathcal{H}_{AB}$, we define a quantity which maps the bipartite state to a nonnegative real number as: 
\begin{eqnarray}
&&\mathcal{Q}_{\rm KD}^{AB}(\varrho_{AB})\nonumber\\
&:=&\inf_{\{\ket{a_1,b_1}\}}\sup_{\{\ket{a_2,b_2}\}}\sum_{a_1,b_1,a_2,b_2}\big|{\rm Im}{\rm Pr}_{\rm KD}(a_1,b_1;a_2,b_2|\varrho_{AB})\big|\nonumber\\
&=&\inf_{\{\ket{a_1,b_1}\}}\sup_{\{\ket{a_2,b_2}\}}\sum_{a_1,b_1,a_2,b_2}\big|{\rm Im}\braket{a_2,b_2|(\Pi_A^{a_1}\otimes\Pi_B^{b_1})\varrho_{AB}|a_2,b_2}\big|\nonumber\\
&=&\inf_{\{\ket{a_1,b_1}\}}\sup_{\{\ket{a_2,b_2}\}}\sum_{a_1,b_1,a_2,b_2}\frac{1}{2}\big|\braket{a_2,b_2|[(\Pi_A^{a_1}\otimes\Pi_B^{b_1}),\varrho_{AB}]|a_2,b_2}\big|,
\label{two-sided KD nonclassical correlation} 
\end{eqnarray}
where the supremum and the infimum are taken over the set $\mathcal{B}_{\rm op}(\mathcal{H}_{AB})$ of all orthonormal product bases of the Hilbert space $\mathcal{H}_{AB}$. 

$\mathcal{Q}_{\rm KD}^{AB}(\varrho_{AB})$ is thus the total sum of the absolute imaginary part of the KD quasiprobability ${\rm Pr}_{\rm KD}(a_1,b_1;a_2,b_2|\varrho_{AB})$ defined in Eq. (\ref{KD quasiprobability 1}), first maximized over all orthonormal product bases $\{\ket{a_2,b_2}\}\in\mathcal{B}_{\rm op}(\mathcal{H}_{AB})$, and then minimized over all orthonormal product bases $\{\ket{a_1,b_1}\}\in\mathcal{B}_{\rm op}(\mathcal{H}_{AB})$.

Note that the first and the second bases in the definition of the KD quasiprobability are not treated symmetrically in the definitions of $\mathcal{Q}_{\rm KD}^A(\varrho_{AB})$ and $\mathcal{Q}_{\rm KD}^{AB}(\varrho_{AB})$. Moreover, we stress that the search for the minimum over the first bases is performed after the maximization over the second bases is done. Conceptually, for a given first orthornormal basis $\{\ket{a_1}\}$ for $\mathcal{Q}_{\rm KD}^A(\varrho_{AB})$ in Eq. (\ref{one-sided KD nonclassical correlation}) (respectively, $\{\ket{a_1,b_1}\}$ for $\mathcal{Q}_{\rm KD}^{AB}(\varrho_{AB})$ in Eq. (\ref{two-sided KD nonclassical correlation})), maximizing over all second bases $\{\ket{a_2,b_2}\}\in\mathcal{B}_{\rm op}(\mathcal{H}_{AB})$ means that we are seeking for the largest noncommutativity between $\{(\Pi_A^{a_1}\otimes\mathbb{I}_B)\}$ (respectively, $\{(\Pi_A^{a_1}\otimes\Pi_B^{b_1})\}$) and $\varrho_{AB}$ under the $l_1$ norm. On the other hand, the minimization over the first local bases $\{\ket{a_1}\}\in\mathcal{B}_{\rm o}(\mathcal{H}_A)$ for $\mathcal{Q}_{\rm KD}^A(\varrho_{AB})$ (respectively, over $\{\ket{a_1,b_1}\}\in\mathcal{B}_{\rm op}(\mathcal{H}_{AB})$ for $\mathcal{Q}_{\rm KD}^{AB}(\varrho_{AB})$) means that we search for those classical-quantum state of Eq. (\ref{classical-quantum state}) (respectively, classical-classical state of Eq. (\ref{classical-classical state})) that is least incompatible with, or in some sense closest to, the bipartite state $\varrho_{AB}$. This goes along with the spirit of the distance-based measure of general quantum correlation in which one computes the minimum distance between the bipartite state and the set of all classical-quantum (respectively, classical-classical) states \cite{Adesso quantum correlation beyond entanglement review}. 

Finally, the above definition can be extended to more than two parties straightforwardly. For example, given a tripartite quantum state $\varrho_{ABC}$ on a finite-dimensional Hilbert space $\mathcal{H}_{ABC}$, we can define the following nonnegative quantity: 
\begin{eqnarray}
&&\mathcal{Q}_{\rm KD}^A(\varrho_{ABC})\nonumber\\
&:=&\inf_{\{\ket{a_1}\}}\sup_{\{\ket{a_2,b_2,c_2}\}}\sum_{a_1,a_2,b_2,c_2}|{\rm Im}{\rm Pr}_{\rm KD}(a_1;a_2,b_2,c_2|\varrho_{ABC})|\nonumber\\
&=&\inf_{\{\ket{a_1}\}}\sup_{\{\ket{a_2,b_2,c_2}\}}\sum_{a_1,a_2,b_2,c_2}|{\rm Im}\braket{a_2,b_2,c_2|(\Pi_A^{a_1}\otimes \mathbb{I}_B\otimes \mathbb{I}_C)\varrho_{ABC}|a_2,b_2,c_2}|. 
\label{one-sided KD nonclassical correlation in tripartite state}
\end{eqnarray}
Here, the generalized KD quasiprobability is defined as ${\rm Pr}_{\rm KD}(a_1;a_2,b_2,c_2)=\sum_{b_1,c_1}{\rm Pr}_{\rm KD}(a_1,b_1,c_1,;a_2,b_2,c_2|\varrho_{ABC}):={\rm Tr}((\Pi_A^{a_2}\otimes\Pi_B^{b_2}\otimes\Pi_C^{c_2})(\Pi_A^{a_1}\otimes\mathbb{I}_B\otimes\mathbb{I}_C)\varrho_{ABC})$, the supremum is taken over the set $\mathcal{B}_{\rm op}(\mathcal{H}_{ABC})$ of all orthonormal product bases $\{\ket{a_2,b_2,c_2}\}$ of $\mathcal{H}_{ABC}$ and the infimum is taken over the set $\mathcal{B}_{\rm o}(\mathcal{H}_A)$ of all orthonormal bases $\{\ket{a_1}\}$ of $\mathcal{H}_A$. 

Consider first $\mathcal{Q}_{\rm KD}^A(\varrho_{AB})$ defined in Eq. (\ref{one-sided KD nonclassical correlation}). We argue that it satisfies the following plausible requirements for a quantifier of general quantum correlation \cite{Adesso quantum correlation beyond entanglement review}. 
\\
{\bf Proposition 1}. {\it Faithfulness}, i.e., $\mathcal{Q}_{\rm KD}^A(\varrho_{AB})$ is vanishing if and only if the bipartite state $\varrho_{AB}$ belongs to the class of classical-quantum state of Eq. (\ref{classical-quantum state}). \\
{\bf Proof}. Let us first suppose that the bipartite state $\varrho_{AB}$ belongs to the class of classical-quantum state, namely there exists an orthonormal basis $\{\ket{k}_A\}$ of $\mathcal{H}_A$ of the subsystem $A$ so that $\varrho_{AB}$ can be expressed as in Eq. (\ref{classical-quantum state}). Then, we can choose $\{\Pi_A^k=\ket{k}\bra{k}_A\}$ as $\{\Pi_A^{a_1}\}$  in the definition of $\mathcal{Q}_{\rm KD}^A(\varrho_{AB})$ in Eq. (\ref{one-sided KD nonclassical correlation}). In this case, since $[(\Pi_A^k\otimes\mathbb{I}_B),\varrho_{AB}]=0$ for all $k$, we have $\mathcal{Q}_{\rm KD}^A(\varrho_{AB})=0$ as per definition. Conversely, suppose $\mathcal{Q}_{\rm KD}^A(\varrho_{AB})=0$. Then, from the definition of $\mathcal{Q}_{\rm KD}^A(\varrho_{AB})$ in Eq. (\ref{one-sided KD nonclassical correlation}), there must be an orthonormal basis $\{\Pi_A^k\}$ of the subsystem $A$ so that $\braket{a_2,b_2|[(\Pi_A^k\otimes\mathbb{I}_B),\varrho_{AB}]|a_2,b_2}=0$ for all $k$ and for all possible choices of the second bases $\{\ket{a_2,b_2}\}\in\mathcal{B}_{\rm op}(\mathcal{H}_{AB})$. This can only be true if $[(\Pi_A^k\otimes\mathbb{I}_B),\varrho_{AB}]=0$ for all $k$. It first implies that $\varrho_{AB}$ is separable. Moreover, taking the trace over $B$, we also get $[\Pi_A^k,\varrho_A]=0$ for all $k$, where $\varrho_A={\rm Tr}_B\varrho_{AB}$, which means that $\{\Pi_A^k\}$ is the complete set of eigenprojectors of $\varrho_{A}$. Hence, we have $\varrho_A=\sum_k p_k\Pi_A^k$ for some $\{p_k\}$, $p_k\ge0$, $\sum_kp_k=1$. Finally, any separable state $\varrho_{AB}$ with the reduced density operator $\varrho_A=\sum_k p_k\Pi_A^k$ must take the form of classical-quantum state of Eq. (\ref{classical-quantum state}). \qed
\\
{\bf Proposition 2}. {\it Invariant under local unitary transformation}, i.e., $\mathcal{Q}_{\rm KD}^A\big((U_A\otimes U_B)\varrho_{AB}(U_A^{\dagger}\otimes U_B^{\dagger})\big)=\mathcal{Q}_{\rm KD}^A(\varrho_{AB})$, where $U_{A(B)}$ is any unitary operator applying locally on subsystem $A(B)$. \\
{\bf Proof}. This comes directly from the definition of $\mathcal{Q}_{\rm KD}^A(\varrho_{AB})$ in Eq. (\ref{one-sided KD nonclassical correlation}) as follows: 
\begin{eqnarray}
&&\mathcal{Q}_{\rm KD}^A\big((U_A\otimes U_B)\varrho_{AB}(U_A^{\dagger}\otimes U_B^{\dagger})\big)\nonumber\\
&=&\inf_{\{\ket{a_1}\}}\sup_{\{\ket{a_2,b_2}\}}\sum_{a_1,a_2,b_2}\big|{\rm Im}\big(\langle a_2,b_2|(U_A\otimes U_B)(U_A^{\dagger}\otimes U_B^{\dagger})\nonumber\\
&&\hspace{20mm}\cdot\hspace{1mm}(\Pi_A^{a_1}\otimes\mathbb{I}_B)(U_A\otimes U_B)\varrho_{AB}(U_A^{\dagger}\otimes U_B^{\dagger})|a_2,b_2\rangle\big)\big|\nonumber\\
&=&\inf_{\{\ket{a_1}\}}\sup_{\{\ket{a'_2,b'_2}\}}\sum_{a_1,a_2',b_2'}\big|{\rm Im}\langle a_2',b_2'|(U_A^{\dagger}\Pi_A^{a_1}U_A\otimes\mathbb{I}_B)\varrho_{AB}|a_2',b_2'\rangle\big|\nonumber\\
&=&\inf_{\{\ket{a'}\}}\sup_{\{\ket{a_2',b_2'}\}}\sum_{a'_1,a_2',b_2'}\big|{\rm Im}\braket{a_2',b_2'|(\Pi_A^{a_1'}\otimes\mathbb{I}_B)\varrho_{AB}|a_2',b_2'}\big|\nonumber\\
&=&\mathcal{Q}_{\rm KD}^A(\varrho_{AB}).
\label{invariant under local unitary}
\end{eqnarray}
Here, we have inserted an identity  $(U_A\otimes U_B)(U_A^{\dagger}\otimes U_B^{\dagger})=\mathbb{I}$ in the first equality. To get the second equality, we have defined a new orthonormal second product basis: $\{\ket{a_2'}\otimes\ket{b_2'}=(U_A^{\dagger}\otimes U_B^{\dagger})\ket{a_2}\otimes\ket{b_2}\}$. Moreover, the third equality is obtained by identifying a new set of first local orthonormal basis as $\{\ket{a'_1}\}=\{U_A^{\dagger}\ket{a_1}\}$. Noting that the above local transformations of bases do not change the set of orthonormal product bases $\mathcal{B}_{\rm op}(\mathcal{H}_{AB})$ and the set of local orthonormal basis $\mathcal{B}_{\rm o}(\mathcal{H}_A)$ over which we respectively perform the maximization and the minimization in Eq. (\ref{one-sided KD nonclassical correlation}), we thus have  $\sup_{\{\ket{a_2,b_2}\}}(\cdot)=\sup_{\{\ket{a_2',b_2'}\}}(\cdot)$ and $\inf_{\{\ket{a_1}\}}(\cdot)=\inf_{\{\ket{a_1'}\}}(\cdot)$. This observation gives the last equality in Eq. (\ref{invariant under local unitary}). We emphasize that the use of product bases for the first and second bases to define the KD quasiprobability, and the maximization and the minimization over these bases, are indispensable to get a quantity which is invariant under any local unitary operation. \qed
\\ 
{\bf Proposition 3}. {\it Monotonicity}: nonincreasing under any local completely positive trace-preserving (CPTP) operation or quantum channel $\Phi_B$ on subsystem $B$, i.e., $\mathcal{Q}_{\rm KD}^A(({\rm id}_A\otimes\Phi_B)\varrho_{AB})\le \mathcal{Q}_{\rm KD}^A(\varrho_{AB})$, where ${\rm id}_A$ denotes an identity superoperator acting on subystem $A$.  \\
{\bf Proof}. First, according to the Stinespring's theorem, any CPTP operation or quantum channel can be implemented by a dilation on a larger Hilbert space, wherein the system is made contact with an ancilla in a state $\varrho_E$ on the Hilbert space $\mathcal{H}_E$, let them interact via some global unitary, and then followed by partial tracing over the ancilla as: $({\rm id}_A\otimes\Phi_B)(\varrho_{AB})={\rm Tr}_E\big((\mathbb{I}_A\otimes U_{BE})(\varrho_{AB}\otimes\varrho_E)(\mathbb{I}_A\otimes U_{BE})^{\dagger}\big)$, where $U_{BE}$ is the unitary interaction applying on the subsystem $B$ and the ancilla $E$. Using this expression, we then have 
\begin{eqnarray}
&&\mathcal{Q}_{\rm KD}^A(({\rm id}_A\otimes\Phi_B)\varrho_{AB})\nonumber\\
&=&\inf_{\{\ket{a_1}\}}\sup_{\{\ket{a_2,b_2}\}}\sum_{a_1,a_2,b_2}\big|{\rm Im}(\langle a_2,b_2|(\Pi_A^{a_1}\otimes\mathbb{I}_B)\nonumber\\
&&\hspace{20mm}\cdot\hspace{1mm}{\rm Tr}_E((\mathbb{I}_A\otimes U_{BE})(\varrho_{AB}\otimes\varrho_E)(\mathbb{I}_A\otimes U_{BE})^{\dagger})|a_2,b_2\rangle)\big|\nonumber\\
&=&\inf_{\{\ket{a_1}\}}\sup_{\{\ket{a_2,b_2}\}}\sum_{a_1,a_2,b_2}\big|{\rm Im}(\sum_{e_2}\langle a_2,b_2,e_2|(\Pi_A^{a_1}\otimes\mathbb{I}_B\otimes\mathbb{I}_E)\nonumber\\
&&\hspace{20mm}\cdot\hspace{1mm}(\mathbb{I}_A\otimes U_{BE})(\varrho_{AB}\otimes\varrho_E)(\mathbb{I}_A\otimes U_{BE})^{\dagger}|a_2,b_2,e_2\rangle)\big|,
\label{proof of monotonicity step 1}
\end{eqnarray}
where we have inserted the basis $\{\ket{e_2}\}$ of the Hilbert space $\mathcal{H}_E$ of the ancilla in the second line. Expanding $U_{BE}^{\dagger}\ket{b_2,e_2}=\sum_{b'_2,e'_2}\braket{b'_2,e'_2|U_{BE}^{\dagger}|b_2,e_2}\ket{b'_2,e'_2}$, we have  
\begin{eqnarray}
&&\mathcal{Q}_{\rm KD}^A(({\rm id}_A\otimes\Phi_B)\varrho_{AB})\nonumber\\
&=&\inf_{\{\ket{a_1}\}}\sup_{\{\ket{a_2,b_2}\}}\sum_{a_1,a_2,b_2}\big|\sum_{b_2',b''_2,e_2,e'_2,e''_2}{\rm Im}\big(\langle a_2,b''_2|(\Pi_A^{a_1}\otimes\mathbb{I}_B)\varrho_{AB}|a_2,b'_2\rangle\nonumber\\
&\times&\braket{e''_2|\varrho_E|e'_2}\braket{b''_2,e''_2|U_{BE}|b_2,e_2}\braket{b_2,e_2|U^{\dagger}_{BE}|b'_2,e'_2}\big)\big|.
\label{proof of monotonicity step 2}
\end{eqnarray}
One can check that for $b'_2\neq b''_2$ and $e'_2\neq e''_2$, each term inside the bracket ${\rm Im}(\cdots)$ has a partner that is its complex conjugate so that their sum are real. Hence, only the terms with $b'_2=b''_2$ and $e'_2=e''_2$ give nonvanishing contribution. In this case, Eq. (\ref{proof of monotonicity step 2}) becomes 
\begin{eqnarray}
&&\mathcal{Q}_{\rm KD}^A(({\rm id}_A\otimes\Phi_B)\varrho_{AB})\nonumber\\
&\le &\inf_{\{\ket{a_1}\}}\sup_{\{\ket{a_2}\}}\sum_{a_1,a_2,b_2'}\big|{\rm Im}\big(\langle a_2,b'_2|(\Pi_A^{a_1}\otimes\mathbb{I}_B)\varrho_{AB}|a_2,b'_2\rangle\big)\big|\sum_{e'_2}\braket{e'_2|\varrho_E|e'_2}\nonumber\\
&\times&\sum_{b_{2*},e_2}\braket{b'_2,e'_2|U_{BE}|b_{2*},e_2}\braket{b_{2*},e_2|U^{\dagger}_{BE}|b'_2,e'_2}\nonumber\\
&\le&\inf_{\{\ket{a_1}\}}\sup_{\{\ket{a_2,b'_2}\}}\sum_{a_1,a_2}\sum_{b_2'}|{\rm Im}\big(\langle a_2,b'_2|(\Pi_A^{a_1}\otimes\mathbb{I}_B)\varrho_{AB}|a_2,b'_2\rangle\big)|\nonumber\\
&=&\mathcal{Q}_{\rm KD}^A(\varrho_{AB}). 
\label{proof of monotonicity step 3}
\end{eqnarray}
Here, $\{\ket{b_{2*}}\}$ is a second basis which achieves the supremum, and we have used the normalization $\sum_{b_{2*},e_2}\ket{b_{2*},e_2}\bra{b_{2*},e_2}=\mathbb{I}_{BE}$, $U_{BE}U_{BE}^{\dagger}=\mathbb{I}_{BE}$, and $\braket{e'_2|\varrho_E|e'_2}$ is real and nonnegative satisfying $\sum_{e'_2}\braket{e'_2|\varrho_E|e'_2}=1$. 
\qed

Hence, $\mathcal{Q}_{\rm KD}^A(\varrho_{AB})$ defined in Eq. (\ref{one-sided KD nonclassical correlation}) indeed satisfies the above requirements expected for a quantifier of general quantum correlation. Accordingly, we shall hereon refer to it as one-sided KD nonclassical correlation. Besides satisfying the above three plausible constraints, the one-sided KD nonclassical correlation also has the following desirable properties. 
\\
{\bf Proposition 4}. Convexity, i.e., $\mathcal{Q}_{\rm KD}^A(\sum_kp_k\varrho_{AB}^k)\le\sum_kp_k\mathcal{Q}_{\rm KD}^A(\varrho_{AB}^k)$, where $\{p_k\}$ are probabilities: $p_k\ge 0$, $\sum_kp_k=1$.
\\
{\bf Proof}. This is due to triangle inequality and the fact that $\{p_k\}$ are real and nonnegative. \qed
\\
{\bf Proposition 5}. Nonincreasing when one or more of the parties are discarded (or, traced over), i.e., $\mathcal{Q}_{\rm KD}^A(\varrho_{ABC})\ge\mathcal{Q}_{\rm KD}^A(\varrho_{AB})$, where $\varrho_{AB}={\rm Tr}_C\varrho_{ABC}$, and equality is reached when the state of $C$ is totally uncorrelated with that of $AB$. \\
{\bf Proof}. One has, from the definition of $\mathcal{Q}_{\rm KD}^A(\varrho_{AB})$ in Eq. (\ref{one-sided KD nonclassical correlation}) and $\mathcal{Q}_{\rm KD}^{A}(\varrho_{ABC})$ in Eq. (\ref{one-sided KD nonclassical correlation in tripartite state}), 
\begin{eqnarray}
&&\mathcal{Q}_{\rm KD}^A(\varrho_{ABC})\nonumber\\
&:=&\inf_{\{\ket{a_1}\}}\sup_{\{\ket{a_2,b_2,c_2}\}}\sum_{a_1,a_2,b_2,c_2}|{\rm Im}\braket{a_2,b_2,c_2|(\Pi_A^{a_1}\otimes \mathbb{I}_B\otimes \mathbb{I}_C)\varrho_{ABC}|a_2,b_2,c_2}|\nonumber\\
&\ge&\inf_{\{\ket{a_1}\}}\sup_{\{\ket{a_2,b_2,c_2}\}}\sum_{a_1,a_2,b_2}|{\rm Im}\sum_{c_2}\braket{a_2,b_2,c_2|(\Pi_A^{a_1}\otimes \mathbb{I}_B\otimes \mathbb{I}_C)\varrho_{ABC}|a_2,b_2,c_2}|\nonumber\\
&=&\inf_{\{\ket{a_1}\}}\sup_{\{\ket{a_2,b_2}\}}\sum_{a_1,a_2,b_2}|{\rm Im}\braket{a_2,b_2|(\Pi_A^{a_1}\otimes\mathbb{I}_B)\varrho_{AB}|a_2,b_2}|\nonumber\\
&=&\mathcal{Q}_{\rm KD}^A(\varrho_{AB}). 
\label{nonincreasing under ignoring one of the parties}
\end{eqnarray}
It is clear that equality is reached when the subsystem $C$ is totally uncorrelated with the rest, i.e., $\mathcal{Q}_{\rm KD}^A(\varrho_{AB}\otimes\varrho_C)=\mathcal{Q}_{\rm KD}^A(\varrho_{AB})$, by virtue of the fact that $\braket{c_2|\varrho_C|c_2}$ is real and nonnegative and $\sum_{c_2}\braket{c_2|\varrho_C|c_2}=1$. Hence, adding or removing an (independent) ancilla does not change the one-sided KD nonclassical correlation as intuitively expected. \qed

Finally, another important requirement for a quantifier to be a bonafide measure of quantum correlation is that for pure states, it should reduce to a measure of entanglement \cite{Adesso quantum correlation beyond entanglement review}. This captures the intuition that for pure states, separability is equivalent to no nonclassical correlation, so that nonclassical correlation for pure state must arise solely from quantum entanglement. We have not yet been able to clarify this important issue. However, in the next section, we prove that for general pure bipartite states of arbitrary finite dimension, the one-sided KD nonclassical correlation defined in Eq. (\ref{one-sided KD nonclassical correlation}) gives a lower bound to the linear entropy of entanglement in the state. Hence, for pure bipartite states, the one-sided KD nonclassical correlation can be seen as a faithful witness of entanglement.  

One can check that $\mathcal{Q}_{\rm KD}^{AB}(\varrho_{AB})$ defined in Eq. (\ref{two-sided KD nonclassical correlation}) also satisfies property (i) of faithfulness, i.e., it is vanishing if and only if the bipartite state $\varrho_{AB}$ takes the form of classical-classical state of Eq. (\ref{classical-classical state}). It is also invariant under local unitary transformation satisfying property (ii). Property (iii) of monotonicity cannot be defined for $\mathcal{Q}_{\rm KD}^{AB}(\varrho_{AB})$ in the bipartite setting. Instead, one can prove for a tripartite state that $\mathcal{Q}_{\rm KD}^{AB}(({\rm id}_{AB}\otimes\Phi_C)\varrho_{ABC})\le \mathcal{Q}_{\rm KD}^{AB}(\varrho_{ABC})$, where $\Phi_C$ is a completely positive trace-preserving operation on $C$. One can also show that $\mathcal{Q}_{\rm KD}^{AB}(\varrho_{AB})$ is convex. Moreover, discarding one or more parties does not increase its value, e.g., we have $\mathcal{Q}_{\rm KD}^{AB}(\varrho_{ABC})\ge\mathcal{Q}_{\rm KD}^{AB}(\varrho_{AB})$, $\varrho_{AB}={\rm Tr}_C\varrho_{ABC}$, with equality when $\varrho_{ABC}=\varrho_{AB}\otimes\varrho_{C}$. Hence, $\mathcal{Q}_{\rm KD}^{AB}(\varrho_{AB})$ defined in Eq. (\ref{two-sided KD nonclassical correlation}) possesses desirable properties for the quantifier of two-sided general quantum correlation, and accordingly, we refer to it as two-sided KD nonclassical correlation.  

Let us further show that, for any bipartite state, the one-sided and the two-sided KD nonclassical correlation sets a lower bound to the negativity of quantumness, a measure of general quantum correlation which quantifies the amount of entanglement that can be activated via a local measurement of the subsystem during the pre-measurement stage \cite{Nakano negativity of quantumness trace norm l1 norm,Piani discord generates entanglement via measurement}. 
\\
{\bf Proposition 6}. For any bipartite state $\varrho_{AB}$ on a finite-dimensional Hilbert space, the following ordering of quantity applies:
\begin{eqnarray}
\mathcal{Q}_{\rm KD}^A(\varrho_{AB})\le\mathcal{Q}_{\rm KD}^{AB}(\varrho_{AB})\le\mathcal{Q}_{l_1}^{AB}(\varrho_{AB}),
\label{KD nonclassical correlation and negativity of quantumness}
\end{eqnarray}
where $\mathcal{Q}_{l_1}^{AB}(\varrho_{AB})$ is a measure of general quantum correlation based on the $l_1$-norm measure of (local) coherence \cite{Adesso quantum correlation beyond entanglement review} which is defined as 
\begin{eqnarray}
\mathcal{Q}_{l_1}^{AB}(\varrho_{AB})&:=&\inf_{\{\ket{a,b}\}}\sum_{a'\neq a,b'\neq b}\big|\braket{a,b|\varrho_{AB}|a',b'}\big|\nonumber\\
&=&\inf_{\{\ket{a,b}\}}\sum_{a,a',b,b'}\big|\braket{a,b|\varrho_{AB}|a',b'}\big|-1. 
\label{general quantum correlation from l1-norm coherence}
\end{eqnarray}
It is equal to twice of the total (two-sided) negativity of quantumness \cite{Nakano negativity of quantumness trace norm l1 norm,Piani discord generates entanglement via measurement}. 
\\
{\bf Proof}. We first show that the one-sided KD nonclassical correlation is always less than or equal to the two-sided KD nonclassical correlation:
\begin{eqnarray}
&&\mathcal{Q}_{\rm KD}^A(\varrho_{AB})\nonumber\\
&=&\inf_{\{\ket{a_1}\}}\sup_{\{\ket{a_2,b_2}\}}\sum_{a_1,a_2,b_2}\big|{\rm Im}\braket{a_2,b_2|(\Pi_A^{a_1}\otimes\mathbb{I}_B)\varrho_{AB}|a_2,b_2}\big|\nonumber\\
&=&\inf_{\{\ket{a_1,b_1}\}}\sup_{\{\ket{a_2,b_2}\}}\sum_{a_1,a_2,b_2}\big|{\rm Im}\braket{a_2,b_2|(\Pi_A^{a_1}\otimes\sum_{b_1}\Pi_B^{b_1})\varrho_{AB}|a_2,b_2}\big|\nonumber\\
&\le&\inf_{\{\ket{a_1,b_1}\}}\sup_{\{\ket{a_2,b_2}\}}\sum_{a_1,b_1,a_2,b_2}\big|{\rm Im}\braket{a_2,b_2|(\Pi_A^{a_1}\otimes\Pi_B^{b_1})\varrho_{AB}|a_2,b_2}\big|\nonumber\\
&=&\mathcal{Q}_{\rm KD}^{AB}(\varrho_{AB}), 
\label{the one-sided is less than or equal the two-sided} 
\end{eqnarray}
where we have noted the fact that the choice of the first basis $\{\Pi_B^{b_1}\}$ which resolves the identity $\mathbb{I}_B$ is arbitrary in the second line, and the last equality is just the definition of  two-sided KD nonclassical correlation of Eq. (\ref{two-sided KD nonclassical correlation}). This relation extends to more than two parties. For example, one has $\mathcal{Q}_{\rm KD}^A(\varrho_{ABC})\le \mathcal{Q}_{\rm KD}^{AB}(\varrho_{ABC})\le \mathcal{Q}_{\rm KD}^{ABC}(\varrho_{ABC})$. 

On the other hand, from the definition of the two-sided KD nonclassical correlation we have 
\begin{eqnarray}
&&\mathcal{Q}_{\rm KD}^{AB}(\varrho_{AB})\nonumber\\
&=&\inf_{\{\ket{a_1,b_1}\}}\sup_{\{\ket{a_2,b_2}\}}\sum_{a_1,b_1,a_2,b_2}\big|{\rm Im}\braket{a_2,b_2|(\Pi_A^{a_1}\otimes\Pi_B^{b_1})\varrho_{AB}|a_2,b_2}\big|\nonumber\\
&\le&\inf_{\{\ket{a_1,b_1}\}}\sup_{\{\ket{a_2,b_2}\}}\sum_{a_2,b_2}\sum_{a_1'\neq a_1,b_1'\neq b_1}\big|\braket{a_1,b_1|\varrho_{AB}|a_1',b_1'}\big|\nonumber\\
&&\times\big|\braket{a_2|a_1}\braket{b_2|b_1}\braket{a_1'|a_2}\braket{b_1'|b_2}\big|\nonumber\\
&=&\inf_{\{\ket{a_1,b_1}\}}\sum_{a_1'\neq a_1,b_1'\neq b_1}\big|\braket{a_1,b_1|\varrho_{AB}|a_1',b_1'}\big|\nonumber\\
&&\times\sum_{a_{2*}}\big|\braket{a_{2*}|a_1}\braket{a_1'|a_{2*}}\big|\sum_{b_{2*}}\big|\braket{b_{2*}|b_1}\braket{b_1'|b_{2*}}\big|,
\label{two-sided KD nonclassical correlation to l1 norm correlation} 
\end{eqnarray}
where we have inserted the completeness relations $\sum_{a_1'}\ket{a_1'}\bra{a_1'}=\mathbb{I}_A$ and $\sum_{b_1'}\ket{b_1'}\bra{b_1'}=\mathbb{I}_B$ for the first orthonormal bases of $A$ and $B$, and $\{\ket{a_{2*}}\}$ and $\{\ket{b_{2*}}\}$ are second orthonormal bases which achieve the supremum. On the other hand, using the Cauchy-Schwartz inequality, we have 
\begin{eqnarray}
&&\sum_{a_{2*}}|\braket{a_{2*}|a_1}| |\braket{a_1'|a_{2*}}|\le\big(\sum_{a_{2*}}|\braket{a_{2*}|a_1}|^2\sum_{a_{2*}'}|\braket{a_1'|a_{2*}'}|^2\big)^{1/2}=1,\nonumber\\
&&\sum_{b_{2*}}|\braket{b_{2*}|b_1}| |\braket{b_1'|b_{2*}}|\le\big(\sum_{b_{2*}}|\braket{b_{2*}|b_1}|^2\sum_{b_{2'*}}|\braket{b_1'|b_{2*}'}|^2\big)^{1/2}=1,
\label{optimal bases and MUB} 
\end{eqnarray}
where we have again used the completeness relation $\sum_{a_{2*}}\ket{a_{2*}}\bra{a_{2*}}=\mathbb{I}_A$ and $\sum_{b_{2*}}\ket{b_{2*}}\bra{b_{2*}}=\mathbb{I}_B$ for the second bases of $A$ and $B$. Given the first basis $\{\ket{a_1(b_1)}\}$, the equalities in Eq. (\ref{optimal bases and MUB}) are attained when the second basis $\{\ket{a_{2*}(b_{2*})}\}$ are (subsystem wise) mutually unbiased with the first basis so that $|\braket{a_{2*}(b_{2*})|a_1(b_1)}|=1/\sqrt{d_{A(B)}}$ for all $a_1(b_1)$ and $a_{2*}(b_{2*})$, where $d_{A(B)}$ is the dimension of the Hilbert space of subsystem $A(B)$. Upon inserting Eq. (\ref{optimal bases and MUB}) into Eq. (\ref{two-sided KD nonclassical correlation to l1 norm correlation}), and noting Eqs. (\ref{general quantum correlation from l1-norm coherence}) and (\ref{the one-sided is less than or equal the two-sided}), we finally obtain Eq. (\ref{KD nonclassical correlation and negativity of quantumness}). 
\qed  

We proceed to give an illustration of concrete computations of the KD nonclassical correlation in a simple bipartite state. We first note that the calculation of the KD nonclassical correlation defined in Eqs. (\ref{one-sided KD nonclassical correlation}) and (\ref{two-sided KD nonclassical correlation}) is in general analytically intractable. It involves optimization over all possible product bases of the Hilbert space of the multipartite system, which in general takes the form of an optimization of multivariable nonlinear function. This analytical difficulty is also suffered by many other measures of general quantum correlation whose computations typically involve optimization over certain class of measurements. Analytical results are known only for low dimensional systems with certain symmetries \cite{Luo analytical two-qubit quantum discord,Ali analytical two-qubit X-state quantum discord}. 

For example, let us compute the one-sided KD nonclassical correlation of maximally entangled two-qubit state of Eq. (\ref{maximally entangled two-qubit state}). For the purpose of computation, we express the second bases for the Hilbert spaces of subsystem $A$ and $B$, i.e., $\{\ket{a_2}\}=\{\ket{a_{2+}},\ket{a_{2-}}\}$ and $\{\ket{b_2}\}=\{\ket{b_{2+}},\ket{b_{2-}}\}$, using the following Bloch sphere parameterization: 
\begin{eqnarray}
\ket{a_{2+}(b_{2+})}&=&\cos\frac{\alpha_{a_2(b_2)}}{2}\ket{0}_{A(B)}+\sin\frac{\alpha_{a_2(b_2)}}{2}e^{i\beta_{a_2(b_2)}}\ket{1}_{A(B)},\nonumber\\
\ket{a_{2-}(b_{2-})}&=&\sin\frac{\alpha_{a_2(b_2)}}{2}\ket{0}_{A(B)}-\cos\frac{\alpha_{a_2(b_2)}}{2}e^{i\beta_{a_2(b_2)}}\ket{1}_{A(B)},
\label{postselection bases}
\end{eqnarray}
where $\alpha_{a_2(b_2)}\in[0,\pi]$, and $\beta_{a_2(b_2)}\in[0,2\pi)$. One can scan over all the second orthonormal product bases for the qubit $A$ and $B$, i.e., all the bases $\{\ket{a_2,b_2}\}\in\mathcal{B}_{\rm op}(\mathcal{H}_{AB})$, by varying the angles $(\alpha_{a_2},\beta_{a_2})$ and $(\alpha_{b_2},\beta_{b_2})$ over their ranges of values. Furthermore, let us parameterize all the possible first local orthonormal bases on subsystem $A$, i.e., $\{\ket{a_1}\}=\{\ket{a_{1+}},\ket{a_{1-}}\}$, as 
\begin{eqnarray}
\ket{a_{1+}}&=&\cos\frac{\theta_{a_1}}{2}\ket{0}_{A}+\sin\frac{\theta_{a_1}}{2}e^{i\eta_{a_1}}\ket{1}_{A},\nonumber\\
\ket{a_{1-}}&=&\sin\frac{\theta_{a_1}}{2}\ket{0}_{A}-\cos\frac{\theta_{a_1}}{2}e^{i\eta_{a_1}}\ket{1}_{A}, 
\label{local measurement basis}
\end{eqnarray}
$\theta_{a_1}\in[0,\pi]$, $\eta_{a_1}\in[0,2\pi)$. Inserting all the above ingredients into the one-sided nonclassical correlation of Eq. (\ref{one-sided KD nonclassical correlation}), we obtain 
\begin{eqnarray}
&&\mathcal{Q}_{\rm KD}^A(\ket{\Psi_{AB}^{\rm me}}\bra{\Psi_{AB}^{\rm me}})\nonumber\\
&=&\inf_{\{\ket{a_1}\}}\sup_{\{\ket{a_2,b_2}\}}\sum_{a_1,a_2,b_2}\big|{\rm Im}\braket{a_2,b_2|(\Pi_A^{a_1}\otimes\mathbb{I}_B)|\Psi_{AB}^{\rm me}}\braket{\Psi_{AB}^{\rm me}|a_2,b_2}\big|\nonumber\\
&=&0.99999\cdots\sim 1. 
\label{one-sided KD nonclassical correlation for two maximally entangled qubits state}
\end{eqnarray}
The sign `$\sim$' denotes that the result is obtained numerically. We show in the next section that the above value in fact maximizes the one-sided KD nonclassical correlation for pure two-qubit state. 

Next, as an example of a separable mixed state with a nonvanishing KD nonclassical correlation, we consider the Werner state for $2\times 2$ dimension \cite{Werner state}: 
\begin{eqnarray}
\varrho_{AB}^{\rm W}=\frac{1-p}{4}(\mathbb{I}_A\otimes\mathbb{I}_B)+p\ket{\Psi_{AB}^{\rm me}}\bra{\Psi_{AB}^{\rm me}}. 
\label{Werner state}
\end{eqnarray}
It is known that the above Werner state is separable for $p<1/3$. The one-sided nonclassical correlation can be straightforwardly computed to get, noting Eq. (\ref{one-sided KD nonclassical correlation for two maximally entangled qubits state}), 
\begin{eqnarray}
\mathcal{Q}_{\rm KD}^A(\varrho^{\rm W}_{AB})=p\mathcal{Q}_{\rm KD}^A(\ket{\Psi_{AB}^{\rm me}}\bra{\Psi_{AB}^{\rm me}})\sim p,
\end{eqnarray}
where the first equality is obtained directly from the definition in Eq. (\ref{one-sided KD nonclassical correlation}). Hence, it is nonvanishing in the regime $p<1/3$ when the Werner state is separable. We note that the above value of the one-sided KD nonclassical correlation for the Werner state is equal to the square root of the geometric discord based on Hilbert-Schmidt distance \cite{Dakic discord and remote state preparation}. Noting Eq. (\ref{the one-sided is less than or equal the two-sided}), for the above Werner state, the two-sided KD nonclassical correlation is thus larger than or equal to $p$.    

Let us make further remark. We have argued in the previous work \cite{Agung KD-nonreality coherence} that the term inside the infimum in Eqs. (\ref{one-sided KD nonclassical correlation}) or (\ref{two-sided KD nonclassical correlation}) can be used as a quantifier of coherence relative to an orthonormal basis. For example, the following quantity:
\begin{eqnarray}
C_{\rm KD}(\varrho_{AB};\{\ket{a_1,b_1}\}):=\sup_{\{\ket{a_2,b_2}\}}\sum_{a_1,b_1,a_2,b_2}\big|{\rm Im}\braket{a_2,b_2|(\Pi_A^{a_1}\otimes\Pi_B^{b_1})\varrho_{AB}|a_2,b_2}\big|,
\label{KD coherence relative to product basis}
\end{eqnarray}
is a faithful quantifier of coherence of $\varrho_{AB}$ relative to the orthonormal product basis $\{\ket{a_1,b_1}\}$, called KD coherence, satisfying certain desirable requirements. The two-sided KD nonclassical correlation of Eq. (\ref{two-sided KD nonclassical correlation}) can thus be expressed as 
\begin{eqnarray}
\mathcal{Q}_{\rm KD}^{AB}(\varrho_{AB})=\inf_{\{\ket{a_1,b_1}\}}C_{\rm KD}(\varrho_{AB};\{\ket{a_1,b_1}\}).
\end{eqnarray} 
Namely, it is the infimum of the KD coherence of the multipartite state $\varrho_{AB}$ relative to all orthonormal product bases $\{\ket{a_1,b_1}\}\in\mathcal{B}_{\rm op}(\mathcal{H}_{AB})$ of the Hilbert space $\mathcal{H}_{AB}$. 

\section{Statistical and operational meaning\label{Statistical and information theoretic meaning}}

\subsection{KD nonclassical correlation as a witness for local quantum uncertainty, pure state entanglement and measurement-induced nonlocality\label{KD nonclassical correlation as the witness for genuine local quantum uncertainty and pure state entanglement}}

Recall that the one-sided and the two-sided KD nonclassical correlations are defined by exploiting the failure of commutativity between the PVM associated with any local basis and the multipartite state. Such noncommutativity is one of the sources of uncertainty in quantum measurement. It is therefore instructive to discuss the relation between the KD nonclassical correlation and the uncertainty arising in the measurement described by the PVM associated with a local basis, over the multipartite state. We have the following proposition.
\\
{\bf Proposition 7}. The one-sided and two-sided KD nonclassical correlations give lower bounds to the minimum uncertainty of the outcomes of all measurements described by local (product) PVM, as follows    
\begin{eqnarray}
\label{KD nonclassical correlation versus local quantum uncertainty - one sided}
\mathcal{Q}_{\rm KD}^A(\varrho_{AB})&\le&\inf_{\{\Pi_A^{a}\}}\sum_{a}\Delta_{(\Pi_A^{a}\otimes\mathbb{I}_B)}(\varrho_{AB}),\\ 
\label{KD nonclassical correlation versus local quantum uncertainty - two sided}
\mathcal{Q}_{\rm KD}^{AB}(\varrho_{AB})&\le&\inf_{\{\Pi_A^{a}\otimes\Pi_B^{b}\}}\sum_{a,b}\Delta_{(\Pi_A^{a}\otimes\Pi_B^{b})}(\varrho_{AB}).
\end{eqnarray} 
where $\Delta^2_{O}[\varrho ]:={\rm Tr}(O^2\varrho)-({\rm Tr}(O\varrho))^2$ is the quantum variance of the Hermitian observable $O$ in the state $\varrho$. 
\\
{\bf Proof}. The proof is given in the Appendix \ref{Proof of w-correlation versus local quantum uncertainty} \qed. 

Hence, the strength of the KD nonclassical correlation is limited by the total sum of the uncertainty in measurement described by the local PVM associated with the first local basis, quantified by the quantum standard deviation of all the elements of the local PVM, minimized over all such local PVMs. As a direct corollary of Proposition 7, the nonclassical correlations captured by $\mathcal{Q}_{\rm KD}^{A}(\varrho_{AB})$ and $\mathcal{Q}_{\rm KD}^{AB}(\varrho_{AB})$ give lower bounds to the quantum uncertainties arising in the measurement described by any local PVM. Namely, from Eqs. (\ref{KD nonclassical correlation versus local quantum uncertainty - one sided}) and (\ref{KD nonclassical correlation versus local quantum uncertainty - two sided}), for any measurement described by local PVM $\{\Pi_A^k\otimes\mathbb{I}_B\}$ and $\{\Pi_A^k\otimes\Pi_B^l\}$, we have 
\begin{eqnarray}
\sum_k\Delta_{(\Pi_A^k\otimes\mathbb{I}_B)}(\varrho_{AB})&\ge&\inf_{\{\Pi_A^{a}\}}\sum_{a}\Delta_{(\Pi_A^{a}\otimes\mathbb{I}_B)}(\varrho_{AB})\ge \mathcal{Q}_{\rm KD}^A(\varrho_{AB}),\\
\sum_{k,l}\Delta_{(\Pi_A^k\otimes\Pi_B^l)}(\varrho_{AB})&\ge&\inf_{\{\Pi_A^{a}\otimes\Pi_B^{b}\}}\sum_{a,b}\Delta_{(\Pi_A^{a}\otimes\Pi_B^{b})}(\varrho_{AB})\ge \mathcal{Q}_{\rm KD}^{AB}(\varrho_{AB}).
\end{eqnarray}

Recall that the right-hand sides of Eqs. (\ref{KD nonclassical correlation versus local quantum uncertainty - one sided}) and (\ref{KD nonclassical correlation versus local quantum uncertainty - two sided}), i.e., the total sum of the quantum standard deviations of all elements of the local PVM over the state $\varrho_{AB}$, also include the uncertainty arising from the classical mixing when the state is not pure. By contrast, the (one-sided and two sided) KD nonclassical correlations quantifies the uncertainties in the measurement of the local bases which arises intrinsically from the noncommutativity between the local PVM associated with the local bases and the multipartite state \cite{Luo's genuine quantum uncertainty1,Luo's genuine quantum uncertainty2,Korzekwa quantum-classical decomposition,Hall quantum-classical decomposition,Agung Budiyono characterization of trace-norm asymmetry in terms of nonreal weak values}. To this end, it is intriguing to compare the KD nonclassical correlations with the quantifier of nonclassical correlation beyond entanglement based on the minimum value of the Wigner-Yanase skew information \cite{Wigner-Yanase skew information} over certain set of local nondegenerate observables proposed in Ref. \cite{Girolami quantum correlation based on WY skew information}. The Wigner-Yanase skew information is also defined based on the noncommutativity between the state and a Hermitian observable and has been argued to quantify the measurement uncertainty arising genuinely from quantum noncommutativity \cite{Luo Wigner-Yanase skew information as genuine quantum uncertainty}. 

Further, evaluating the optimization on the right-hand side of Eq. (\ref{KD nonclassical correlation versus local quantum uncertainty - one sided}) we obtain the following result connecting the one-sided KD nonclassical correlation to a form of quantum  entropy. 
\\
{\bf Proposition 8}. The one-sided KD nonclassical correlation $\mathcal{Q}_{\rm KD}^A(\varrho_{AB})$ of a bipartite state $\varrho_{AB}$ is upper bounded by the linear entropy of the reduced density operator of the subsystem as 
\begin{eqnarray}
\mathcal{Q}_{\rm KD}^A(\varrho_{AB})\le \sqrt{d_A\big(1-{\rm Tr}\varrho_A^2\big)}, 
\label{general quantum correlation as the lower bound for linear entropy of entanglement}
\end{eqnarray} 
where $\varrho_A={\rm Tr}_B\{\varrho_{AB}\}$ and $d_A$ is the dimension of the subsystem $A$. 
\\
{\bf Proof}. First, noting that ${\rm Tr}((\Pi_A^{a}\otimes\mathbb{I}_B)^2\varrho_{AB})={\rm Tr}((\Pi_A^{a}\otimes\mathbb{I}_B)\varrho_{AB})={\rm Pr}(a|\varrho_A)$ is just the probability to get the outcome $a$ in the measurement described by a PVM $\{\Pi_A^{a}\}$ over the state  $\varrho_A={\rm Tr}_B(\varrho_{AB})$ of the subsystem $A$, and inserting into Eq. (\ref{KD nonclassical correlation versus local quantum uncertainty - one sided}), we obtain 
\begin{eqnarray}
\label{upper bound of one-sided KD nonclassical correlation 0}
\mathcal{Q}_{\rm KD}^A(\varrho_{AB})&\le&\inf_{\{\Pi_A^{a}\}}\sum_{a}\big({\rm Pr}(a|\varrho_A)-{\rm Pr}(a|\varrho_A)^2)^{1/2}\\
&\le&\inf_{\{\Pi_A^{a}\}}d_A^{1/2}\big(1-\sum_{a}{\rm Pr}(a|\varrho_A)^2)^{1/2}\nonumber\\
\label{upper bound of one-sided KD nonclassical correlation}
&=&\inf_{\{\Pi_A^{a}\}}d_A^{1/2}S_2^{1/2}(\{{\rm Pr}(a|\varrho_A)\}),
\end{eqnarray}
where we have used the Jensen inequality and the normalization $\sum_{a}{\rm Pr}(a|\varrho_A)=1$ to get the second line, and 
\begin{eqnarray}
S_2(\{{\rm Pr}(a|\varrho_A)\}):= 1-\sum_{a}{\rm Pr}(a|\varrho_A)^2
\label{Tsallis entropy with index entropy 2}
\end{eqnarray}
is the Tsallis entropy with the entropy index $2$ of the measurement outcome $a$ with a probability ${\rm Pr}(a|\varrho_A)$. 

Let us further show that the minimum of the Tsallis entropy on the right-hand side of Eq. (\ref{upper bound of one-sided KD nonclassical correlation}) is obtained when the PVM $\{\Pi_A^{a}\}$ is just given by the complete set of the eigenprojectors of the reduced density operator $\varrho_A$. First, assume that $\varrho_A$ has the following spectral decomposition: $\varrho_A=\sum_j\lambda_j\Pi^{\lambda_j}_A$, where $\{\Pi^{\lambda_j}_A=\ket{\lambda_j}\bra{\lambda_j}_A\}$ is the set of the eigenprojectors of $\varrho_A$, and $\{\lambda_j\}$, $\lambda_j\ge 0$, $\sum_j\lambda_j=1$, is the associated set of eigenvalues. We thus have ${\rm Pr}(a|\varrho_A)={\rm Tr}(\Pi_A^{a}\varrho_A)=\sum_j\lambda_j{\rm Tr}(\Pi_A^{\lambda_j}\Pi_A^{a})$. Inserting this into the definition of the Tsallis entropy in Eq. (\ref{Tsallis entropy with index entropy 2}), and noting the convexity of the quadratic function, we get, using the Jensen inequality,
\begin{eqnarray}
S_2(\{{\rm Pr}(a|\varrho_A)\})&=&1-\sum_{a}\big(\sum_j\lambda_j{\rm Tr}(\Pi_A^{\lambda_j}\Pi_A^{a})\big)^2\nonumber\\
&\ge&1-\sum_{a}\sum_j\lambda_j^2{\rm Tr}(\Pi_A^{\lambda_j}\Pi_A^{a})\nonumber\\
&=&1-\sum_j\lambda_j^2=S_2(\{\lambda_j\}),
\label{minimum Tsallis entropy}
\end{eqnarray}
where we have used the normalization $\sum_{a}{\rm Tr}(\Pi_A^{\lambda_j}\Pi_A^{a})=\braket{\lambda_j|\lambda_j}=1$. Upon comparing Eq. (\ref{Tsallis entropy with index entropy 2}) to Eq. (\ref{minimum Tsallis entropy}), one finds that the equality in Eq. (\ref{minimum Tsallis entropy}), namely, the minimum of $S_2(\{{\rm Pr}(a|\varrho_A)\})$, is reached when ${\rm Pr}(a|\varrho_A)=\lambda_a={\rm Tr}(\Pi_A^{\lambda_a}\varrho_A)$, i.e., when $\Pi_A^{a}=\Pi_A^{\lambda_a}$ for all $a$, as claimed:
\begin{eqnarray}
\inf_{\{\Pi_A^{a}\}}S_2(\{{\rm Pr}(a|\varrho_A)\})=S_2(\{\lambda_j\}). 
\label{minimum Tsallis entropy 1}
\end{eqnarray} 
Next, inserting Eq. (\ref{minimum Tsallis entropy 1}) into Eq. (\ref{upper bound of one-sided KD nonclassical correlation}), we therefore have
\begin{eqnarray}
\mathcal{Q}_{\rm KD}^A(\varrho_{AB})&\le& d_A^{1/2}S_2^{1/2}(\{\lambda_j\})\nonumber\\
&=& d_A^{1/2}\big(1-\sum_j\lambda_j^2\big)^{1/2}\nonumber\\
&=& d_A^{1/2}\big(1-{\rm Tr}\varrho_A^2\big)^{1/2}.
\label{upper bound of one-sided KD nonclassical correlation optimal}
\end{eqnarray} \qed

Notice that the Tsallis entropy over the eigenvalues of $\varrho_A$, i.e., $S_2(\{\lambda_j\})=1-{\rm Tr}(\varrho_A^2)$, is just the quantum linear entropy associated with state $\varrho_A=\sum_j\lambda_j\Pi_A^{\lambda_j}$. It quantifies the mixedness or impurity of the reduced density operator $\varrho_A={\rm Tr}_B\varrho_{AB}$. Moreover, recall that for a pure bipartite state $\ket{\psi_{AB}}$, the linear entropy of the reduced density matrix $\varrho_A={\rm Tr}_B(\ket{\psi_{AB}}\bra{\psi_{AB}})$ or $\varrho_B={\rm Tr}_A(\ket{\psi_{AB}}\bra{\psi_{AB}})$ can be used as a measure of entanglement in $\ket{\psi_{AB}}$ \cite{Vidal measure of pure state entanglement}. In fact, the upper bound in Eq. (\ref{general quantum correlation as the lower bound for linear entropy of entanglement}), i.e., $\sqrt{d_A(1-{\rm Tr}\varrho_A^2)}$, is proportional to the concurrence for pure state which is given by $C:=\sqrt{2(1-{\rm Tr}\varrho_A^2)}$, from which one can compute the entanglement of formation for two qubits. Noting this, Eq. (\ref{general quantum correlation as the lower bound for linear entropy of entanglement}) therefore shows that for general pure bipartite states $\ket{\psi_{AB}}$, the one-sided KD nonclassical correlation $\mathcal{Q}_{\rm KD}^A(\ket{\psi_{AB}}\bra{\psi_{AB}})$ can be seen as a witness for the quantum entanglement in $\ket{\psi_{AB}}$, namely, it gives a lower bound to the scaled square root of the linear entropy of entanglement  as  
\begin{eqnarray}
\mathcal{Q}_{\rm KD}^A(\ket{\psi_{AB}}\bra{\psi_{AB}})\le d_A^{1/2}(1-{\rm Tr}(\varrho_A^2))^{1/2}, 
\label{KD nonclassical correlation as a witness for linear entropy of entanglement}
\end{eqnarray}
with $\varrho_A={\rm Tr}_B(\ket{\psi_{AB}}\bra{\psi_{AB}})$. Notice further that for pure bipartite states with Schmidt decomposition: $\ket{\psi_{AB}}=\sum_i\sqrt{\lambda_j}\ket{\psi_i}_A\ket{\phi_i}_B$, the measurement-induced nonlocality  \cite{Luo measurement induced nonlocality} yields 
\begin{eqnarray}
N(\ket{\psi_{AB}}\bra{\psi_{AB}})&=&1-\sum_i\lambda_i^2\nonumber\\
&\ge& \mathcal{Q}_{\rm KD}^A(\ket{\psi_{AB}}\bra{\psi_{AB}})^2/d_A, 
\label{measurement-induced nonlocality vs KD nonclassical correlation}
\end{eqnarray}
where the inequality is due to Eq. (\ref{KD nonclassical correlation as a witness for linear entropy of entanglement}). This is also the case for the geometric discord \cite{Dakic necessary and sufficient condition for geometric quantum discord}. Hence, for pure states, the KD nonclassical correlation also gives a lower bound to the measurement induced nonlocality and geometric discord as in Eq. (\ref{measurement-induced nonlocality vs KD nonclassical correlation}).  

As an example, let us evaluate the upper bound in Eq. (\ref{KD nonclassical correlation as a witness for linear entropy of entanglement}) for a bipartite system $AB$ wherein the subsystem $A$ is a qubit, hence, $d_A=2$. Denoting the Schmidt coefficients as $\sqrt{\lambda_+}$ and $\sqrt{\lambda_-}$, with $\lambda_++\lambda_-=1$, we thus have $\varrho_A=\lambda_+\ket{\lambda_+}\bra{\lambda_+}+\lambda_-\ket{\lambda_-}\bra{\lambda_-}$. In this case, the upper bound in Eq. (\ref{KD nonclassical correlation as a witness for linear entropy of entanglement}) is just given by the entanglement concurrence, i.e., $d_A^{1/2}(1-{\rm Tr}(\varrho_A^2))^{1/2}=\sqrt{2}\sqrt{1-(\lambda_+^2+\lambda_-^2)}=2\sqrt{\lambda_+\lambda_-}$. This is maximized and equal to one for a maximally entangled state, i.e., when $\lambda_+=\lambda_-=1/2$. On the other hand, if the subsystem $B$ is also a qubit, as shown numerically in Eq. (\ref{one-sided KD nonclassical correlation for two maximally entangled qubits state}), we have for the maximally entangled state $\mathcal{Q}_{\rm KD}^A(\ket{\psi_{AB}}\bra{\psi_{AB}})=1$. Hence, for maximally entangled two-qubit state, the inequality in Eq. (\ref{KD nonclassical correlation as a witness for linear entropy of entanglement}) becomes equality, i.e., the KD nonclassical correlation takes its maximum value. 

\subsection{Measurement of KD nonclassical correlation and its information theoretical meaning\label{Observation of KD nonclassical correlation and its information theoretical meaning}}

We show in this subsection that KD nonclassical correlation of an unknown quantum state can in principle be measured or estimated directly in experiment, i.e., without recoursing to quantum state tomography. Let us discuss the estimation of the two-sided KD nonclassical correlation. The estimation of the one-sided KD nonclassical correlation follows the same general scheme. First, we write the two-sided KD nonclassical correlation of Eq. (\ref{two-sided KD nonclassical correlation}) as
\begin{eqnarray}
\mathcal{Q}_{\rm KD}^{AB}(\varrho_{AB})&=&\inf_{\{\ket{a_1,b_1}\}}\sup_{\{\ket{a_2,b_2}\}}\sum_{a_1,b_1,a_2,b_2}\Big|{\rm Im}\Big(\frac{\braket{a_2,b_2|(\Pi_A^{a_1}\otimes\Pi_B^{b_1})\varrho_{AB}|a_2,b_2}}{\braket{a_2,b_2|\varrho_{AB}|a_2,b_2}}\Big)\Big|\nonumber\\
&\times&\braket{a_2,b_2|\varrho_{AB}|a_2,b_2}.  
\label{KD nonclassical correlation as conditional weak value}
\end{eqnarray} 
One then notices that the term on the right-hand side of the first line, i.e., 
\begin{eqnarray}
\pi_{a_1b_1}^{\rm w}(\varrho_{AB}|a_2,b_2):=\frac{\braket{a_2,b_2|(\Pi_A^{a_1}\otimes\Pi_B^{b_1})\varrho_{AB}|a_2,b_2}}{\braket{a_2,b_2|\varrho_{AB}|a_2,b_2}}, 
\label{KD quasiprobability as weak value}
\end{eqnarray}
is just the weak value of $\Pi_A^{a_1}\otimes\Pi_B^{b_1}$ with the preselected state $\varrho_{AB}$ and a postselected state $\ket{a_2,b_2}$ \cite{Aharonov weak value,Aharonov-Daniel book,Wiseman weak value,Tamir weak value review}. The KD nonclassical correlation can thus be obtained by averaging the absolute imaginary part of the weak value $\pi_{a_1b_1}^{\rm w}(\varrho_{AB}|a_2,b_2)$ over the probability ${\rm Pr}(a_2,b_2|\varrho_{AB})=\braket{a_2,b_2|\varrho_{AB}|a_2,b_2}$ to get $(a_2,b_2)$, and then followed by the maximization over all the orthonormal product basis $\ket{a_2,b_2}\in\mathcal{B}_{\rm op}(\mathcal{H}_{AB})$ and minimization over the orthonormal product basis $\ket{a_1,b_1}\in\mathcal{B}_{\rm op}(\mathcal{H}_{AB})$.   

We thus have the following general scheme for the measurement of the KD nonclassical correlation. First, we need to be able to scan all the orthonormal product bases of the Hilbert space $\mathcal{H}_{AB}$ of the bipartite system $AB$. Namely, we have to be able to prepare the local orthonormal basis $\{\ket{a(\vec{\lambda}_A)}\}$, where $\vec{\lambda}_A$ is a set of scalar parameters so that their variation scan all the orthonormal bases $\mathcal{B}_{\rm o}(\mathcal{H}_A)$ of the Hilbert space $\mathcal{H}_A$. Similarly, we have to be able to prepare the local orthonormal basis $\{\ket{b(\vec{\lambda}_B)}\}$, so that the variation of $\vec{\lambda}_B$ scans all the orthonormal bases $\mathcal{B}_{\rm o}(\mathcal{H}_B)$ of the Hilbert space $\mathcal{H}_B$. An example of such parameterization of the local orthonormal bases for two-qubit system in Bloch sphere is given in Eqs. (\ref{postselection bases}) and (\ref{local measurement basis}). This means that we have to be able to implement parameteized unitary circuits $U_{\vec{\lambda}_{A_j}}$ and $U_{\vec{\lambda}_{B_j}}$, $j=1,2$, which can transform the standard basis into any other orthonormal bases of the Hilbert space of the subsystem $A$ and $B$. Here, the subscript $j=1,2$ refers to the fact that for each subsystem, we need to be able to prepare two orthonormal bases independently. Equipped with such parameterized unitary circuits, we first pick a value for the set of parameters $(\vec{\lambda}_{A_1},\vec{\lambda}_{B_1},\vec{\lambda}_{A_2},\vec{\lambda}_{B_2})$ to estimate the imaginary part of the weak value $\pi_{a_1(\vec{\lambda}_{A_1})b_1(\vec{\lambda}_{B_1})}^{\rm w}\big(\varrho_{AB}|a_2(\vec{\lambda}_{A_2}),b_2(\vec{\lambda}_{B_2})\big)$ defined in Eq. (\ref{KD quasiprobability as weak value}) via a number of methods suggested in the literatures \cite{Aharonov weak value,Wiseman weak value,Johansen quantum state from successive projective measurement,Johansen weak value from a sequence of strong measurement,Salvail direct measurement KD distribution,Bamber measurement of KD distribution,Lundeen measurement of KD distribution,Thekkadath measurement of density matrix,Lostaglio KD quasiprobability and quantum fluctuation,Jozsa complex weak value,Haapasalo generalized weak value,Cohen estimating of weak value with strong measurements,Vallone strong measurement to reconstruct quantum wave function,Wagner measuring weak values and KD quasiprobability}. Then, we average over the probability ${\rm Pr}(a_2(\vec{\lambda}_{A_2}),b_2(\vec{\lambda}_{B_2})|\varrho_{AB})=\braket{a_2(\vec{\lambda}_{A_2}),b_2(\vec{\lambda}_{B_2})|\varrho_{AB}|a_2(\vec{\lambda}_{A_2}),b_2(\vec{\lambda}_{B_2})}$, vary the parameters $(\vec{\lambda}_{A_2},\vec{\lambda}_{B_2})$ until we get the converging supremum over all the second orthonormal product basis $\{\ket{a_2(\vec{\lambda}_{A_2}),b_2(\vec{\lambda}_{B_2})}\}\in\mathcal{B}_{\rm op}(\mathcal{H}_{AB})$. Finally, we vary the parameters $(\vec{\lambda}_{A_1},\vec{\lambda}_{B_1})$ for the first orthonormal product bases, and repeat the above procedure, until we get the converging infimum value over all the first orthonormal product bases $\{\ket{a_1(\vec{\lambda}_{A_1}),b_1(\vec{\lambda}_{B_1})}\}\in\mathcal{B}_{\rm op}(\mathcal{H}_{AB})$. Hence, we have a hybrid quantum-classical variational circuit \cite{Cerezo VQA review} for computing the KD nonclassical correlation which may be implemented on the near-term quantum hardware \cite{Preskill NISQ era quantum computing}. Let us mention that a different approach of using variational quantum circuit to compute general quantum correlation is suggested in Ref. \cite{Mahdian variational quantum circuit for quantum discord}.    
 
Note that, as argued in Refs. \cite{Johansen weak value best estimation,Hall prior information}, the variance of the imaginary part of the weak value in Eq. (\ref{KD quasiprobability as weak value}) can be seen as the mean squared error of the optimal estimation of the product basis $\{\Pi_A^{a_1}\otimes\Pi_B^{b_1}\}$ based on the outcome of measurement of product basis $\{\Pi_A^{a_2}\otimes\Pi_B^{b_2}\}$. In this sense, Eq. (\ref{KD nonclassical correlation as conditional weak value}), may be interpreted as the minimum mean absolute error of such optimal estimation in the worst case scenario, i.e., the best of the worst. Moreover, the imaginary part of the weak value in Eq. (\ref{KD quasiprobability as weak value}) can also be interpreted as the disturbance of the state $\varrho_{AB}$ due to a unitary translation generated by local Hermitian observable $\Pi_A^{a_1}\otimes\Pi_B^{b_1}$ \cite{Dressel imaginary weak value and disturbance}. Noting this, Eq. (\ref{KD nonclassical correlation as conditional weak value}) can also be interpreted as the minimum of average of such disturbance in the worst case scenario. 

The above scheme for the observation of the two-sided KD nonclassical correlation using weak value measurement and classical optimization can be applied to estimate the one-sided KD nonclassical correlation of Eq. (\ref{one-sided KD nonclassical correlation}) entailing similar statistical meaning. 

\section{Summary and Remarks\label{Summary and Remarks}}

In this paper we discussed a fundamental question: how is the general quantum correlation in a bipartite state, wherein entanglement is a subset, encoded in the associated KD quasiprobability? More specifically, how is the nonclassicality captured by the general quantum correlation in a bipartite state related to the nonclassical values of the KD quasiprobability? We showed that the sum of the absolute imaginary part of the KD quasiprobability defined over a pair of orthonormal product bases, suitably optimized over all such bases, can be used to quantify the general quantum correlation in the associated bipartite state, satisfying certain desirable requirements. We discussed the relation between the quantifier of general quantum correlation, called KD nonclassical correlation, with the negativity of quantumness, minimum local quantum uncertainty, and entanglement and measurement-induced nonlocality for pure states. We also gave a general scheme for the estimation of the KD nonclassical correlation using a hybrid quantum-classical variational circuit implementable in the near-term quantum hardware. Our results suggest intriguing fundamental links between the quantum-classical division and contrast captured by general nonclassical correlation, and the concept of nonclassicality captured by the nonclassical values of the Kirkwood-Dirac quasiprobability and the associated strange weak values in the multipartite setting. 

\begin{acknowledgments}  
\end{acknowledgments} 

\appendix 

\section{Proof of Proposition 7 \label{Proof of w-correlation versus local quantum uncertainty}}

First, we have, from Eq. (\ref{one-sided KD nonclassical correlation}), 
\begin{eqnarray}
&&\mathcal{Q}_{\rm KD}^A(\varrho_{AB})\nonumber\\
&=&\inf_{\{\ket{a_1}\}}\sup_{\{\ket{a_2,b_2}\}}\sum_{a_1}\sum_{a_2,b_2}\Big|{\rm Im}\Big(\frac{{\rm Tr}(\Pi_{AB}^{a_2b_2}(\Pi_A^{a_1}\otimes\mathbb{I}_B)\varrho_{AB})}{{\rm Tr}(\Pi_{AB}^{a_2b_2}\varrho_{AB})}\Big)\Big|{\rm Tr}(\Pi_{AB}^{a_2b_2}\varrho_{AB})\nonumber\\
&\le&\inf_{\{\ket{a_1}\}}\sup_{\{\ket{a_2,b_2}\}}\sum_{a_1}\Big(\sum_{a_2,b_2}\Big|{\rm Im}\Big(\frac{{\rm Tr}(\Pi_{AB}^{a_2b_2}(\Pi_A^{a_1}\otimes\mathbb{I}_B)\varrho_{AB})}{{\rm Tr}(\Pi_{AB}^{a_2b_2}\varrho_{AB})}\Big)\Big|^2{\rm Tr}(\Pi_{AB}^{a_2b_2}\varrho_{AB})\Big)^{1/2}\nonumber\\
&=&\sum_{a_{1*}}\Big(\sum_{a_{2*},b_{2*}}\Big(\Big|\frac{{\rm Tr}(\Pi_{AB}^{a_{2*}b_{2*}}(\Pi_A^{a_{1*}}\otimes\mathbb{I}_B)\varrho_{AB})}{{\rm Tr}(\Pi_{AB}^{a_{2*}b_{2*}}\varrho_{AB})}\Big|^2\nonumber\\
&&-{\rm Re}\Big(\frac{{\rm Tr}(\Pi_{AB}^{a_{2*}b_{2*}}(\Pi_A^{a_{1*}}\otimes\mathbb{I}_B)\varrho_{AB})}{{\rm Tr}(\Pi_{AB}^{a_{2*}b_{2*}}\varrho_{AB})}\Big)^2\Big){\rm Tr}(\Pi_{AB}^{a_{2*}b_{2*}}\varrho_{AB})\Big)^{1/2}\nonumber\\
&\le&\sum_{a_{1*}}\Big(\sum_{a_{2*},b_{2*}}\frac{|{\rm Tr}(\Pi_{AB}^{a_{2*}b_{2*}}(\Pi_A^{a_{1*}}\otimes\mathbb{I}_B)\varrho_{AB})|^2}{{\rm Tr}(\Pi_{AB}^{a_{2*}b_{2*}}\varrho_{AB})}-\big(\sum_{a_{2*},b_{2*}}{\rm Re}\big({\rm Tr}(\Pi_{AB}^{a_{2*}b_{2*}}(\Pi_A^{a_{1*}}\otimes\mathbb{I}_B)\varrho_{AB})\big)\big)^2\Big)^{1/2}.
\label{from local weak measurement to local quantum uncertainty 0}
\end{eqnarray}
Here $\Pi_{AB}^{a_2b_2}=\Pi_A^{a_2}\otimes\Pi_B^{b_2}$, $\{\ket{a_{2*},b_{2*}}\}$ is a second orthonormal product basis which achieves the supremum, and $\{\Pi_A^{a_{1*}}\}$ is a first orthonormal basis which achieves the infimum. Moreover, we have applied the Jensen inequality to get the third and fifth lines. Next, applying the Cauchy-Schwartz inequality for bounded operators, i.e., $|{\rm Tr}(XY^{\dagger})|^2\le{\rm Tr}(XX^{\dagger}){\rm Tr}(YY^{\dagger})$, to the numerator in the first term on the right-hand side of Eq. (\ref{from local weak measurement to local quantum uncertainty 0}), i.e., 
\begin{eqnarray}
\big|{\rm Tr}(\Pi_{AB}^{a_{2*}b_{2*}}(\Pi_A^{a_{1*}}\otimes\mathbb{I}_B)\varrho_{AB})\big|^2&=&|{\rm Tr}(((\Pi_{AB}^{a_{2*}b_{2*}})^{1/2}(\Pi_A^{a_{1*}}\otimes\mathbb{I}_B)\varrho_{AB}^{1/2})(\varrho_{AB}^{1/2}(\Pi_{AB}^{a_{2*}b_{2*}})^{1/2}))|^2\nonumber\\
&\le&{\rm Tr}(\Pi_{AB}^{a_{2*}b_{2*}}(\Pi_A^{a_{1*}}\otimes\mathbb{I}_B)\varrho_{AB}(\Pi_A^{a_{1*}}\otimes\mathbb{I}_B)){\rm Tr}(\varrho_{AB}\Pi_{AB}^{a_{2*}b_{2*}}), 
\end{eqnarray}
and using the completeness relation $\sum_{a_{2*}b_{2*}}\Pi_{AB}^{a_{2*}b_{2*}}=\mathbb{I}$, we finally obtain Eq. (\ref{KD nonclassical correlation versus local quantum uncertainty - one sided}), i.e.,
\begin{eqnarray}
\mathcal{Q}_{\rm KD}^A(\varrho_{AB})&\le&\sum_{a_*}\big({\rm Tr}((\Pi_A^{a_*}\otimes\mathbb{I}_B)^2\varrho_{AB})-{\rm Tr}((\Pi_A^{a_*}\otimes\mathbb{I}_B)\varrho_{AB})^2)^{1/2}\nonumber\\
&=&\inf_{\{\Pi_A^{a}\}}\sum_{a}\Delta_{(\Pi_A^{a}\otimes\mathbb{I}_B)}(\varrho_{AB}). 
\label{one-sided KD nonclassical correlation is upper bounded by minimum local uncertainty Appendix}
\end{eqnarray}

By following exactly the same steps as above, we also obtain the inequality of Eq. (\ref{KD nonclassical correlation versus local quantum uncertainty - two sided}) for the two-sided KD nonclassical correlation:
\begin{eqnarray}
\mathcal{Q}_{\rm KD}^{AB}(\varrho_{AB}))&\le&\sum_{a_*,b_*}\big({\rm Tr}((\Pi_A^{a_*}\otimes\Pi_B^{b_*})^2\varrho_{AB})-{\rm Tr}((\Pi_A^{a_*}\otimes\Pi_B^{b_*})\varrho_{AB})^2)^{1/2}\nonumber\\
&=&\inf_{\{\Pi_A^{a}\otimes\Pi_B^{b}\}}\sum_{a,b}\Delta_{(\Pi_A^{a}\otimes\Pi_B^{b})}(\varrho_{AB}), 
\label{two-sided KD nonclassical correlation is upper bounded by minimum local uncertainty Appendix}
\end{eqnarray}
where $\Pi_A^{a_*}\otimes\Pi_B^{b_*}=\ket{a_*,b_*}\bra{a_*,b_*}$, with $\{\ket{a_*,b_*}\}$ is a first orthonormal product basis of $\mathcal{H}_{AB}$ which achieves the infimum in Eq. (\ref{two-sided KD nonclassical correlation}). \qed

\end{document}